\documentclass{LMCS}

\def\dOi{10(3:14)2014}
\lmcsheading%
{\dOi}
{1--48}
{}
{}
{Dec.\phantom05, 2012}
{Sep.~03, 2014}
{}

\ACMCCS{[{\bf Software and its engineering}]: Software notations and
  tools---General programming languages---Language features---Data
  types and structures; [{\bf Theory of computation}]: Semantics and
  reasoning---Program semantics---Categorical semantics}


\usepackage{hyperref}
\usepackage{url}
\usepackage{amssymb,amsmath,amsthm}
\usepackage{proof}
\usepackage{stmaryrd}
\usepackage{verbatim} 
\usepackage{mdwlist}

\usepackage[all]{xy}

\newcommand{\opc}{\mathrm{op}}

\newcommand{\id}{\mathsf{id}}    
\newcommand{\comp}{\circ}

\newcommand{\Id}{\mathsf{Id}}    
\newcommand{\Comp}{\cdot}

\newcommand{\cid}{\mathsf{id}^\mathrm{c}}  
\newcommand{\ccomp}{\mathrel{\circ^\mathrm{c}}}

\newcommand{\cId}{\mathsf{Id}^\mathrm{c}}  
\newcommand{\cComp}{\mathrel{\cdot^\mathrm{c}}}

\newcommand{\ee}{\mathsf{e}}
\newcommand{\mm}{\mathsf{m}}

\newcommand{\zt}{{\ast}}

\newcommand{\fst}{\mathsf{fst}}
\newcommand{\snd}{\mathsf{snd}}

\newcommand{\inl}{\mathsf{inl}}
\newcommand{\inr}{\mathsf{inr}}

\newcommand{\Set}{\mathsf{Set}}
\newcommand{\SetB}{\mathbf{Set}}

\newcommand{\Cont}{\mathbf{Cont}}	
\newcommand{\SFun}{[\SetB,\SetB]}
\newcommand{\DCont}{\mathbf{DCont}}
\newcommand{\Comonad}[1]{\mathbf{Comonads({#1})}}
\newcommand{\Comonoid}[1]{\mathbf{Comonoids({#1})}}
\newcommand{\ia}[1]{\{#1\}}
\newcommand{\Monad}[1]{\mathbf{Monads({#1})}}
\newcommand{\Monoid}[1]{\mathbf{Monoids({#1})}}

\newcommand{\csem}[1]{\llbracket #1 \rrbracket^\mathrm{c}}
\newcommand{\cquo}[1]{\ulcorner #1 \urcorner^\mathrm{c}}
\newcommand{\dcsem}[1]{\llbracket #1 \rrbracket^\mathrm{dc}}
\newcommand{\dcquo}[1]{\ulcorner #1 \urcorner^\mathrm{dc}}
\newcommand{\ccosem}[1]{\langle\!\langle #1 \rangle\!\rangle^\mathrm{c}}
\newcommand{\dccosem}[1]{\langle\!\langle #1 \rangle\!\rangle^\mathrm{dc}}

\newcommand{\eps}{\varepsilon}
\newcommand{\de}{\delta}

\newcommand{\depl}{\delta^+} 

\newcommand{\carrierplao}{\overline{D^+_0}}

\newcommand{\depla}{\delta^+_0} 

\newcommand{\carrierplbo}{\overline{D^+_1}}

\newcommand{\deplb}{\delta^+_1} 

\newcommand{\heps}{h^{\varepsilon}}
\newcommand{\teps}{t^{\varepsilon}}
\newcommand{\qeps}{q^{\varepsilon}}
\newcommand{\hde}{h^{\delta}}
\newcommand{\tde}{t^{\delta}}
\newcommand{\qde}{q^{\delta}}

\newcommand{\hepsp}{h^{\varepsilon'}}
\newcommand{\tepsp}{t^{\varepsilon'}}
\newcommand{\qepsp}{q^{\varepsilon'}}
\newcommand{\hdep}{h^{\delta'}}
\newcommand{\tdep}{t^{\delta'}}
\newcommand{\qdep}{q^{\delta'}}

\newcommand{\dn}{\mathrel{\downarrow}}
\renewcommand{\o}{\mathsf{o}}
\newcommand{\pl}{\mathrel{\oplus}}

\newcommand{\dnt}{\mathrel{\downarrow}}  
\newcommand{\ot}{\mathsf{o}}
\newcommand{\plt}{\mathrel{\oplus}}

\newcommand{\dnpl}{\mathrel{\downarrow^+}}
\newcommand{\plpl}{\mathrel{\oplus^+}}

\newcommand{\dntpl}{\mathrel{\downarrow^+}} 
\newcommand{\pltpl}{\mathrel{\oplus^+}}

\newcommand{\dnpll}{\mathrel{\overline{\dn_0^+}}}
\newcommand{\dnplr}{\mathrel{\overline{\dn_1^+}}}

\newcommand{\plpll}{\mathrel{\overline{\oplus_0^+}}}
\newcommand{\plplr}{\mathrel{\overline{\oplus_1^+}}}

\newcommand{\posplao}{\overline{P_0^+}}
\newcommand{\shapeao}{\overline{S_0}}
\newcommand{\dnpla}{\mathrel{\downarrow_0^+}}
\newcommand{\plpla}{\mathrel{\oplus_0^+}}

\newcommand{\posplbo}{\overline{P_1^+}}
\newcommand{\shapebo}{\overline{S_1}}
\newcommand{\dnplb}{\mathrel{\downarrow_1^+}}
\newcommand{\plplb}{\mathrel{\oplus_1^+}}

\newcommand{\dnp}{\mathrel{\downarrow'}}
\newcommand{\op}{\mathsf{o'}}
\newcommand{\plp}{\mathrel{\oplus'}}

\newcommand{\dna}{\mathrel{\downarrow_0}}
\newcommand{\oa}{\mathsf{o}_0}
\newcommand{\pla}{\mathrel{\oplus_0}}

\newcommand{\dnb}{\mathrel{\downarrow_1}}
\newcommand{\ob}{\mathsf{o}_1}
\newcommand{\plb}{\mathrel{\oplus_1}}

\newcommand{\pii}{\pi}

\newcommand{\sh}{\mathsf{s}}

\newcommand{\Nat}{\mathsf{Nat}}
\newcommand{\Fin}{\mathsf{Fin}}

\newcommand{\Maybe}{\mathsf{Maybe}}
\newcommand{\just}{\mathsf{just}}
\newcommand{\nothing}{\mathsf{nothing}}

\newcommand{\List}{\mathsf{List}}

\newcommand{\Str}{\mathsf{Str}}

\newcommand{\tfzero}{t^{f_0}}
\newcommand{\tfone}{t^{f_1}}
\newcommand{\tfbarzero}{\overline{\tfzero}}
\newcommand{\tfbarone}{\overline{\tfone}}
\newcommand{\qfzero}{q^{f_0}}
\newcommand{\qfone}{q^{f_1}}
\newcommand{\qfbarzero}{\overline{\qfzero}}
\newcommand{\qfbarone}{\overline{\qfone}}
\newcommand{\Pplusbarzero}{\overline{P_0^+}}
\newcommand{\Pplusbarone}{\overline{P_1^+}}

\begin{document}

\title[When is a Container a Comonad?]{When is a Container a
  Comonad?\rsuper *}

\author[D.~Ahman]{Danel Ahman\rsuper a}
\address{{\lsuper a}%
Laboratory for Foundations of Computer Science, 
School of Informatics, 
University of Edinburgh,
10 Crichton Street,
Edinburgh EH8 9AB, United Kingdom}
\email{d.ahman@ed.ac.uk}
\thanks{{\lsuper{a,b,c}}The authors were supported by the ERDF funded CoE project EXCS,
  the Estonian Ministry of Education and Research target-financed
  theme no.\ 0140007s12, and the Estonian Science
  Foundation grants no.\ 9219 and 9475.}

\author[J.~Chapman]{James Chapman\rsuper b}
\address{{\lsuper{b,c}}%
Institute of Cybernetics at Tallinn University of Technology,
Akadeemia tee 21, 12618 Tallinn, Estonia}
\email{\{james,tarmo\}@cs.ioc.ee}

\author[T.~Uustalu]{Tarmo Uustalu\rsuper c}
\address{\vspace{-18 pt}}


\titlecomment{{\lsuper *}This article is a revised and expanded version of the FoSSaCS 2012 conference paper~\cite{ACU}.}

\begin{abstract}
  Abbott, Altenkirch, Ghani and others have taught us that many
  parameterized datatypes (set functors) can be usefully analyzed via
  container representations in terms of a set of shapes and a set of
  positions in each shape. This paper builds on the observation that
  datatypes often carry additional structure that containers alone do
  not account for. We introduce directed containers to capture the
  common situation where every position in a data-structure determines
  another data-structure, informally, the sub-data-structure rooted by
  that position.  Some natural examples are non-empty lists and
  node-labelled trees, and data-structures with a designated position
  (zippers). While containers denote set functors via a
  fully-faithful functor, directed containers interpret
  fully-faithfully into comonads. But more is true: every comonad
  whose underlying functor is a container is represented by a directed
  container. In fact, directed containers are the same as containers
  that are comonads.  We also describe some constructions of directed
  containers. We have formalized our development in the dependently
  typed programming language Agda.
\end{abstract}
\maketitle


\section{Introduction}
\label{sec:intro}

Containers, as introduced by Abbott, Altenkirch and Ghani
\cite{Abbott2005}
are a neat representation for a wide class of parameterized datatypes
(set functors) in terms of a set of shapes and a set of positions in
each shape. They cover lists, colists, streams, various kinds of
trees, etc. Containers can be used as a ``syntax'' for programming
with these datatypes and reasoning about them, as can the strictly
positive datatypes and polynomial functors of Dybjer~\cite{dybjer1997}, 
Moerdijk and Palmgren~\cite{moerdijk.palmgren},
Gambino and Hyland~\cite{gambino.hyland:poly}, and
Kock~\cite{kock:polyfuncandtrees}.  The theory of this class of
datatypes is elegant, as they are well-behaved in many respects.

This paper proceeds from the observation that datatypes often carry
additional structure that containers alone do not account for. We
introduce directed containers to capture the common situation in
programming where every position in a data-structure determines another
data-structure, informally, the sub-data-structure rooted by that
position. Some natural examples of such data-structures are non-empty 
lists and node-labelled trees, and data-structures with a designated 
position or focus (zippers). In the former case, the sub-data-structure 
is a sublist or a subtree.  In the latter case, it is the whole 
data-structure but with the focus moved to the given position.

We show that directed containers are no less neat than containers.
While containers denote set functors via a fully-faithful functor,
directed containers interpret fully-faithfully into comonads. They
admit some of the constructions that containers do, but not others:
for instance, two directed containers cannot be composed in general.
Our main result is that every comonad whose underlying functor is the
interpretation of a container is the interpretation of a directed
container. So the answer to the question in the title of this paper
is: a container is a comonad exactly when it is a directed container.
In more precise terms, the category of directed containers is the
pullback of the forgetful functor from the category of comonads to
that of set functors along the interpretation functor of containers.
This also means that a directed container is the same as a comonoid
in the category of containers.

In the core of the paper, we study directed containers on
$\SetB$. Toward the end of the paper we point it out that the
development could also be carried out more generally in locally Cartesian closed categories (LCCCs) and yet
more generally in categories with pullbacks.

In our mathematics, we use syntax similar to the dependently typed
functional programming language Agda \cite{norell:thesis,agda}. If some
function argument will be derivable in most contexts, we mark it as
implicit by enclosing it/its type in braces in the function's type
declaration and either give this argument in braces or omit it in the
definition and applications of the function.

We have formalized the central parts of the theory presented in
Agda. The development is available at
\url{http://cs.ioc.ee/~danel/dcont.html}.

\subsection*{Structure of the Article}

In Section~\ref{sec:containers}, we review the basic theory of
containers, showing also some examples. We introduce containers and
their interpretation into set functors. We show some constructions of
containers such as the coproduct of containers. In
Section~\ref{sec:dcontainers}, we revisit our examples and introduce
directed containers as a specialization of containers and describe
their interpretation into comonads. Our main result, that a container
is a comonad exactly when it is directed, is the subject of
Section~\ref{sec:pullback}. In Section~\ref{sec:constructions}, we
look at some constructions, in particular the cofree directed
container and the focussed container (zipper) construction. In
addition, we also introduce strict directed containers and construct
the product of two strict directed containers in the category of
directed containers. Intuitively, a strict directed container is a
directed container where no position in a non-root subshape of a shape
translates to its root.  In Section~\ref{sec:monads}, we ask whether a
similar characterization is possible for containers that are monads
and hint that this is the case. In Section~\ref{sec:cointerp}, we show
that interpreting the opposite of the category of directed containers
into set functors gives monads. In Section~\ref{sec:polycom}, we hint
how the directed container theory (presented in the paper for $\SetB$)
could be developed in the more general setting of categories with
pullbacks. We briefly summarize related work in
Section~\ref{sec:related} and conclude with outlining some directions
for future work in Section~\ref{sec:concl}. The proofs of the main
results of Sections~\ref{sec:dcontainers} and \ref{sec:constructions}
appear in Appendices A and B.

We spend a section on the background theory of containers as they are
central for our paper but relatively little known, but assume that the
reader knows about comonads, monoidal categories and comonoids.

\subsection*{Differences from the FoSSaCS 2012 Conference Version} 

This article is a revised and expanded version of the FoSSaCS 2012
conference paper \cite{ACU}. We have added many of the proofs that
were omitted from the conference version. We have rearranged
the different constructions on directed containers into a separate
section, namely Section~\ref{sec:constructions}.  In
Section~\ref{sec:cofree}, we give a detailed discussion of cofree
directed containers. In Section~\ref{sec:products},
which is entirely new, we define strict directed containers and coideal
comonads and give an explicit formula for the product of two strict
directed containers.

Likewise entirely new are the sections on cointerpreting directed
containers in Section \ref{sec:cointerp} and directed containers in
categories with pullbacks in Section \ref{sec:polycom}.


\section{Containers}
\label{sec:containers}

We begin with a recap of containers. We introduce the category of
containers and the fully-faithful functor into the category of set
functors defining the interpretation of containers and show that these
are monoidal. We also recall some basic constructions of
containers. For proofs of the propositions in this section and further
information, we refer the reader to Abbott et
al.~\cite{Abbott2005,abbott:phd}.

\subsection{Containers}

Containers are a form of ``syntax'' for datatypes. A \emph{container}
$S \lhd P$ is given by a set $S : \Set$ of \emph{shapes} and a
shape-indexed family $P : S \to \Set$ of \emph{positions}. Intuitively, 
shapes are ``templates'' for data-structures and positions
identify ``blanks'' in these templates that can be filled with data.

\begin{exa}
The datatype of lists is represented by $S \lhd P$ where the shapes $S
= \Nat$ are the possible lengths of lists and the positions $P\, s =
\Fin \, s = \{0,\ldots,s-1\}$ provide $s$ places for data in lists of
length $s$.  Non-empty lists are obtained by letting $S = \Nat$ and
$P\, s = \Fin \, (s+1)$ (so that shape $s$ has $s+1$ rather than $s$
positions). 
\end{exa}

\begin{exa}
Streams are characterized by a single shape with natural number
positions: $S = 1 = \{\zt\}$ and $P\, \zt = \Nat$. The singleton
datatype has one shape and one position: $S = 1$, $P\, \zt = 1$.
\end{exa}

A \emph{morphism} between containers $S \lhd P$ and $S' \lhd P'$ is a
pair $t \lhd q$ of maps $t : S \to S'$ and $q : \Pi \ia{s : S}.\, P'\,
(t\, s) \to P\, s$ (the shape map and position map).  
Note how the positions are mapped backwards.
The intuition is that, if a function between two datatypes does not
look at the data, then the shape of a data-structure given to it must
determine the shape of the data-structure returned and the data in any
position in the shape returned must come from a definite position in
the given shape. 

\begin{exas}
\mbox{}
\begin{itemize}
\item The head function, sending a non-empty
list to a single data item, is determined by the maps $t : \Nat \to 1$
and $q : \Pi \ia{s : \Nat}.\, 1 \to \Fin\, (s+1)$ defined by $t\, \_ =
\zt$ and $q\, \zt = 0$.

\item The tail function, sending a non-empty list to
a list, is represented by $t : \Nat \to \Nat$ and $q : \Pi \ia{s :
  \Nat}.\, \Fin\, s \to \Fin\, (s+1)$ defined by $t\, s = s $ and $q\,
p = p + 1$.

\item For the function dropping every second element of a
non-empty list, the shape and position maps $t : \Nat \to \Nat$ and $q
: \Pi \ia{s : \Nat}.\, \Fin\, (s \div 2 + 1) \to \Fin\, (s + 1)$ are
$t\, s = s \div 2$ and $q\, p = p * 2$.

\item For self-append of a non-empty list, they are $t : \Nat \to
  \Nat$ and $q : \Pi \ia{s : \Nat}.\, \Fin\, (s * 2 + 2) \to \Fin\, (s +
  1)$ defined by $t\, s = s * 2 + 1$ and $q\, \{s\}\, p = p \mod (s + 1)$.

\item For reversal of
non-empty lists, they are $t : \Nat \to \Nat$ and $q : \Pi \ia{s :
  \Nat}.\, \Fin\, (s + 1) \to \Fin\, (s + 1)$ defined by $t\, s = s$
and $q\, \{s\}\, p = s - p$. 
\end{itemize}
(See Prince et al.~\cite{Prince2008} for more similar examples.)
\end{exas}

The \emph{identity} morphism $\cid \{C\}$ on a container $C = S \lhd
P$ is defined by $\cid = \id\, \ia{S} \lhd \lambda \ia{s}.\, \id\,
\ia{P\, s}$. The \emph{composition} $h \ccomp h'$ of container
morphisms $h = t \lhd q$ and $h' = t' \lhd q'$ is defined by $h \ccomp
h' = t \comp t' \lhd \lambda \ia{s}.\, q'\, \ia{s} \comp q\, \ia{t'\,
  s}$. Composition of container morphisms is associative,  
identity is the unit.

\begin{prop}
Containers form a category $\Cont$.
\end{prop}

\subsection{Interpretation of Containers}

To map containers into datatypes made of data-structures that have the
positions in some shape filled with data, we must equip containers 
with a ``semantics''.

For a container $C = S \lhd P$, we define its \emph{interpretation}
$\csem{C} : \Set \to \Set$ on sets by $\csem{C}\, X = \Sigma s: S.\,
P\, s \to X$, so that $\csem{C}\, X$ consists of pairs of a shape and
an assignment of an element of $X$ to each of the positions in this
shape, reflecting the intuitive reading that shapes are ``templates" for 
datatypes and positions identify ``blanks" in these templates that can 
be filled in with data. The interpretation
$\csem{C} : \forall \ia{X}, \ia{Y}.\, (X \to Y) \to (\Sigma s: S.\,
P\, s \to X) \to \Sigma s: S.\, P\, s \to Y$ of $C$ on functions is
defined by $\csem{C}\, f\, (s, v) = (s, f \comp v)$. It is
straightforward that $\csem{C}$ preserves identity and composition of
functions, so it is a set functor (as any datatype should be).

\begin{exa}
Our example containers denote the datatypes intended. 
If we let $C$ be
the container of lists, we have $\csem{C}\, X = \Sigma s : \Nat.\,
\Fin\, s \to X \cong \List\, X$. The container of streams interprets
into $\Sigma \zt : 1.\, \Nat \to X \cong \Nat \to X \cong \Str\, X$.
Etc.
\end{exa}

A morphism $h = t \lhd q$ between containers $C = S \lhd P$ and $C =
S' \lhd P'$ is interpreted as a natural transformation between
$\csem{C}$ and $\csem{C'}$, i.e., as a polymorphic function $\csem{h}
: \forall \ia{X}.\, (\Sigma s: S.\, P\, s \to X) \to \Sigma s' : S'.\,
P'\, s' \to X$ that is natural. It is defined by $\csem{h}\, (s, v) =
(t\, s, v \comp q\, \ia{s})$. $\csem{-}$ preserves the identities and
composition of container morphisms.

\begin{exa}
The interpretation of the container morphism $h$ for the
list head function is $\csem{h} : \forall\ia{X}.\, (\Sigma s: \Nat.\,
\Fin\, (s+1) \to X) \to \Sigma \zt: 1.\, 1 \to X$ defined by
$\csem{h}\, (s, v) = (\zt, \lambda \zt.\, v\, 0)$.
\end{exa}

\begin{prop}
$\csem{-}$ is a functor from $\Cont$ to $\SFun$.
\end{prop}

Every natural transformation between container interpretations is the
interpretation of some container morphism. For containers $C = S \lhd
P$ and $C' = S' \lhd P'$, a natural transformation $\tau$ between
$\csem{C}$ and $\csem{C'}$, i.e., a polymorphic function $\tau :
\forall \ia{X}.\, (\Sigma s : S.\, P\, s \to X) \to \Sigma s':
S'.\,P'\, s' \to X$ that is natural, can be ``quoted'' to a container
morphism $\cquo{\tau} = t \lhd q$ between $C$ and $C'$ where $t : S \to
S'$ and $q : \Pi \ia{s : S}.\, P'\, (t\, s) \to P\, s$ are defined by
$\cquo{\tau} = (\lambda s.\, \fst\, (\tau\, \ia{P\, s}\, (s, \id)))
\lhd (\lambda \ia{s}.\, \snd\, (\tau\, \ia{P\, s}\, (s, \id)))$.

For any container morphism $h$, $\cquo{\csem{h}} = h$, and, for any
natural transformation $\tau$ and $\tau'$ between container
interpretations, $\cquo{\tau} = \cquo{\tau'}$ implies $\tau = \tau'$.

\begin{prop}
\label{prop:csemfullyfaithful}
$\csem{-}$ is fully faithful.
\end{prop}

\subsection{Monoidal Structure}

We have already seen the \emph{identity} container
$\cId = 1 \lhd \lambda \zt.\, 1$. The \emph{composition} $C_0 \cComp
C_1$ of containers $C_0 = S_0 \lhd P_0$ and $C_1 = S_1 \lhd P_1$ is the
container $S \lhd P$ defined by $S = \Sigma s : S_0.\, P_0\, s \to
S_1$ and $P\, (s, v) = \Sigma p_0 : P_0\, s.\, P_1\, (v\, p_0)$. It
has as shapes pairs of an outer shape $s$ and an assignment of an
inner shape to every position in $s$.  The positions in the composite
container are pairs of a position $p$ in the outer shape and a
position in the inner shape assigned to $p$. The (horizontal)
composition $h_0 \cComp h_1$ of container morphisms $h _0 = t_0 \lhd
q_0$ and $h_1 = t_1 \lhd q_1$ is the container morphism $t \lhd q$
defined by $t\, (s, v) = (t_0\, s, t_1 \comp v \comp q_0\, \ia{s})$
and $q\, \ia{s, v}\, (p_0, p_1) = (q_0\, \ia{s}\, p_0, q_1\, \ia{v\,
  (q_0\, \ia{s}\, p_0)}\, p_1)$. The horizontal composition preserves
the identity container morphisms and the (vertical) composition of
container morphisms, which means that ${-} \cComp {-}$ is a bifunctor.

$\Cont$ has isomorphisms $\rho : \forall \ia{C}.\, C \cComp \cId \to
C$, $\lambda : \forall \ia{C}.\, \cId \cComp C \to C$ and $\alpha :
\forall \ia{C,C',C''}.\, (C \cComp C') \cComp C'' \to C
\cComp (C' \cComp C'')$, given by 
$\rho = \lambda (s, v).\, s \lhd
\lambda \ia{s,v}.\, \lambda p.\, (p, \zt)$, 
$\lambda = \lambda
(\zt,v).\, v\, \zt \lhd \lambda \ia{\zt,v}.\, \lambda p.\, (\zt, p)$ and 
$\alpha = \lambda ((s, v), v').\, (s , \lambda p.\, (v\, p, \lambda
p'.\, v'\, (p,p'))) \lhd \linebreak \lambda \ia{(s, v), v'}.\, \lambda
(p,(p',p'')).\, ((p, p'), p'')$.
They satisfy Mac Lane's coherence conditions.

\begin{prop}
\label{prop:monoidalcategory}
The category $\Cont$ is a monoidal category.
\end{prop}

There are also natural isomorphisms $\ee : \Id \to \csem{\cId}$ and
$\mm : \forall \ia{C_0,C_1}.\, \csem{C_0} \Comp
\csem{C_1}$ $\to$ $\csem{C_0 \cComp C_1}$ defined by 
$\ee\, x = (\zt,\lambda \zt.\, x)$ and
$\mm\, (s,v) = ((s,\lambda p.\, \fst\,
(v\, p)), \lambda (p,p').\, \snd\, (v\, p)\, p')$
satisfying the appropriate coherence conditions.

\begin{prop}
  The functor $\csem{-}$ is a monoidal functor.
\end{prop}

\subsection{Constructions of Containers}

Containers are closed under various constructions such as products,
coproducts and constant exponentiation, preserved by
interpretation. 

\paragraph{Products} For two containers $C_0 = S_0 \lhd P_0$ and $C_1 = S_1 \lhd
  P_1$, their \emph{product} $C_0 \times C_1$ is the container $S \lhd
  P$ defined by $S = S_0 \times S_1$ and $P\, (s_0, s_1) = P_0\, s_0 +
  P_1\, s_1$. It holds that $\csem{C_0 \times C_1} \cong \csem{C_0}
  \times \csem{C_1}$.
\paragraph{Coproducts} The \emph{coproduct} $C_0 + C_1$ of containers $C_0 = S_0 \lhd
  P_0$ and $C_1 = S_1 \lhd P_1$ is the container $S \lhd P$ defined by
  $S = S_0 + S_1$, $P\, (\inl\, s) = P_0\, s$ and $P\, (\inr\, s) =
  P_1\, s$.  It is the case that $\csem{C_0 + C_1} \cong \csem{C_0} +
  \csem{C_1}$.
\paragraph{Exponentials} For a set $K \in \Set$ and a container $C_0 = S_0 \lhd P_0$, the
  \emph{exponential} $K \to C_0$ is the container $S \lhd P$ where $S
  = K \to S_0$ and $P\, f = \Sigma k : K.\, P \,(f\, k)$. We have that
  $\csem{K \to C_0} \cong K \to \csem{C_0}$.


\section{Directed Containers}
\label{sec:dcontainers}

We now proceed to our contribution, directed containers. We define the
category of directed containers and a fully-faithful functor
interpreting directed containers as comonads, and discuss some
examples and constructions.

\subsection{Directed Containers}
\label{sec:dcont}

Parametrized datatypes often carry some additional structure that is worth making
explicit. For example, each node in a list or non-empty list defines a
sublist (a suffix). In container terms, this corresponds to every
position in a shape determining another shape, the subshape
corresponding to this position. The theory of containers alone does
not account for such additional structure. Directed
containers, studied in the rest of this paper, axiomatize subshapes and
translation of positions in a subshape into the global shape.

\begin{defi}
A \emph{directed container} is a container $S \lhd P$ together with
three operations
\begin{itemize*}
\item ${\dn} : \Pi s : S.\, P\, s \to S$ (the subshape corresponding to a position), 
\item $\o : \Pi \ia{s : S}.\, P\, s$ (the root), 
\item ${\pl} : \Pi \ia{s : S}.\, \Pi p : P\, s.\, P\, (s \dn p) \to P\,
  s$ (translation of subshape positions into positions in the global shape).
\end{itemize*}
satisfying the following two shape equations and three position equations:
\begin{enumerate*}
\item $\forall \ia{s}.\, s \dn \o = s$,
\item $\forall \ia{s, p, p'}.\,  s \dn (p \pl p') = (s \dn p) \dn p'$,
\item $\forall \ia{s, p}.\, p \pl \ia{s}\, \o = p$, 
\item $\forall \ia{s, p}.\, \o\, \ia{s} \pl p = p$, 
\item $\forall \ia{s, p, p', p''}.\, (p \pl \ia{s}\, p') \pl p'' = p \pl (p' \pl p'')$.
\end{enumerate*}
\end{defi}
(Using $\pl$ as an infix operation, we write the first, implicit,
argument next to the operation symbol when we want to give it
explicitly.) Modulo the fact that the positions involved come from
different sets, laws 3--5 are the laws of a monoid. In the special
case $S = 1$, we have exactly one set of positions, namely $P\, \ast$,
and that is a monoid. If $S$ is general, but $s \dn p$ does not depend
on $p$ (in this case $s \dn p = s$ thanks to law 1), then each $P\, s$ is a monoid. (One might also notice that laws
1--2 bear similarity to the laws of a monoid action. If none of $P\,
s$, $\o\, \ia{s}$, $p \oplus\, \ia{s}\, p'$ depends on $s$, then we have
one single monoid and $\dn$ is then a right action of that monoid on
$S$.)

To help explain the operations and laws, we sketch in
Fig.~\ref{fig:dcontainer} a data-structure with nested
sub-data-structures.

\begin{figure}[h]
\[
\xymatrix@C=1.45em@R=1em{
& & & & & & & & \bullet \ar@(l,u)@{->}[]^(0.25){\o \ia{s}} \ar@{-}[ddddddddllllllll] \ar@{-}[ddddddddrrrrrrrr]_>>>>>>{s = s \dn \o\ia{s}} \ar@{->}[dd]^(0.6){p = p \pl \o \ia{s'} = \o \ia{s} \pl p : P s} \ar@{->}@/_5pc/[dddd]_{p \pl p'}  \ar@{->}@/_8.5pc/[ddddddd]_{(p \pl p') \pl p'' = p \pl (p' \pl p'')} \\
& & & & & & & & & \\
& & & & & & & &  \bullet \ar@(l,u)@{.>}[]^(0.25){\o \ia{s'}}  \ar@{.>}[dd]^(0.6){p' : P s'} \ar@{.}[ddddddllllll] \ar@{.}[ddddddrrrrrr]_>>>>>>{s' = s \dn p} \ar@{.>}@/^4.5pc/[ddddd]^(0.4){p' \pl p''} & &  \\
& & & & & & & & & & & \\
& & & & & & & & \bullet \ar@{-->}[ddd]^(0.65){p'' : P s''} \ar@{--}[ddddllll]^>>>>>>{s'' = s \dn (p \pl p') = s' \dn p'} \ar@{--}[ddddrrrr]& & & &  \\
& & & & & & & & & & & & & \\
& & & & & & & & & & & & & \\
& & & & & & & & \bullet & & & & & & \\
\ar@{-}[rrrrrrrrrrrrrrrr] & & & &  & & & & & & & & & & & &
}
\]
\caption{A data-structure with two nested sub-data-structures}
\label{fig:dcontainer}
\end{figure}

The global shape $s$ is marked with a solid boundary and has a
root position $\o\, \ia{s}$. Then, any position $p$ in $s$
determines a shape $s' = s \dn p$, marked with a dotted boundary, to
be thought of as the subshape of $s$ given by this position. The root
position in $s'$ is $\o\, \ia{s'}$. Law 3 says that its translation $p
\pl \o\, \ia{s'}$ into a position in shape $s$ is $p$, reflecting the
idea that the subshape given by a position should have that position
as the root.

By law 1, the subshape $s \dn \o\, \ia{s}$ corresponding to the root
position $\o \ia{s}$ in the global shape $s$ is $s$ itself. Law 4,
which is only well-typed thanks to law 1, stipulates that the
translation of position $p$ in $s \dn \o\, \ia{s}$ into a position in
$s$ is just $p$ (which is possible, as $P\, (s \dn \o\, \ia{s}) = P\, s$).

A further position $p'$ in $s'$ determines a shape $s'' = s'
\dn p'$. But $p'$ also translates into a position $p \pl p'$ in $s$
and that determines a shape $s \dn (p \pl p')$. Law 2 says that $s''$ and 
$s \dn (p \pl p')$ are the same shape, which is marked by a dashed boundary 
in the figure. Finally, law 5 (well-typed only because of law 2) 
says that the two alternative ways to translate a position $p''$ in shape
$s''$ into a position in shape $s$ agree with each other.

\begin{exa}
Lists cannot form a directed container, as the shape $0$ (for the
empty list), having no positions, has no possible root position. 

But the container of \emph{non-empty lists} (with $S = \Nat$ and $P\,s
= \Fin\,(s +1)$) is a directed container with respect to
\emph{non-empty suffixes} as sublists. The subshape given
by a position $p$ in a shape $s$ (for lists of length $s+1$) is the
shape of the corresponding suffix, given by $s \dn p = s - p$. The
root $\o\, \ia{s}$ is the position $0$ of the head node. A position in
the global shape is recovered from a position $p'$ in the subshape of
the position $p$ by $p \pl p' = p + p'$.

Fig. \ref{fig:dcontainernelist} shows an example of the shape and 
positions of a non-empty list with length 6, i.e., with shape $s = 5$.
This figure also shows that
the subshape determined by a position $p = 2$ in the global shape $s$
is $s' = s \dn p = 5 - 2 = 3$ and a position $p' = 1$ in $s'$ is
rendered as the position $p \pl p' = 2 + 1 = 3$ in the initial shape.
\begin{figure}[t]
\small
\[
\xymatrix@C=5em@R=1.2em{
& & & & & \\
& & & & & \\
\bullet \ar@{->}@/_0.8pc/[rr]_(0.8){p = 2} \ar@{->}@/^0.8pc/[rrr]^{p \pl p' = 2 + 1 = 3} \ar@{-}[uurr]_>{s = 5} \ar@{-}[ddrr]  & \bullet & \bullet \ar@{.>}@/_0.8pc/[r]_>{p' = 1} \ar@{.}[uurr]_>{s' = s \dn p = 5 - 2 = 3} \ar@{.}[ddrr] & \bullet & \bullet & \bullet \\
& & & & & \\
& & & & & 
}
\]
\caption{The shape and positions for non-empty lists of length 6}
\label{fig:dcontainernelist}
\end{figure}
Clearly one could also choose prefixes as subshapes and the
last node of a non-empty list as the root, but this gives an isomorphic
directed container. 
\end{exa}

\begin{exa}
Non-empty lists also give rise to an entirely
different directed container structure that has \emph{cyclic shifts}
as ``sublists'' (this example was suggested to us by Jeremy Gibbons).
The subshape at each position is the global shape ($s \dn p = s$). The
root is still $\o\, \ia{s} = 0$. The interesting part is that
translation into the global shape of a subshape position is defined by
$p \pl \ia{s}\, p' = (p + p') \mod (s + 1)$, satisfying all the required laws.
\end{exa}

\begin{figure}
\small
\[
\xymatrix@C=4.8em@R=1.2em{
& & & & & \\
& & & & & \\
\bullet \ar@{->}@/_1pc/[rr]_(0.8){p=2} \ar@{->}@/^1pc/[rrrr]^{p \pl p' = 2 + \, 2 = 4} \ar@{-}[uurr]_>{s = \zt} \ar@{-}[ddrr] & \bullet  & \bullet \ar@{.>}@/_1pc/[rr]_(0.8){p' = 2} \ar@{.}[uurr]_>{s' = s \dn p = \zt} \ar@{.}[ddrr] & \bullet & \bullet & \bullet & ... \\
& & & & & \\
& & & & &
}
\]
\normalsize
\caption{The shape and positions for streams}
\label{fig:dcontainerstream}
\end{figure}

\begin{exa}
The container of \emph{streams} ($S = 1$, $P\, \zt = \Nat$) carries a
  very trivial directed container structure given by $\zt \dn p =
  \zt$, $\o = 0$ and $p \pl p' = p +
  p'$. Fig.~\ref{fig:dcontainerstream} shows how a position $p = 2$ in
  the only possible global shape $s = \zt$ and a position $p' = 2$ in
  the equal subshape $s' = s \dn p = \zt$ give back a position $p + p'
  = 4$ in the global shape.

  This directed container is nothing else than the monoid $(\Nat,
  0, +)$ seen as a directed container.
\end{exa}

Similarly to the theory of containers, one can also define morphisms
between directed containers. 
\begin{defi}
A \emph{morphism} between two directed
containers $(S \lhd P, \dn, \o, \pl)$ and \linebreak $(S' \lhd P', {\dnp}, \op,
{\plp})$ is a morphism $t \lhd q$ between the containers $S \lhd P$
and $S' \lhd P'$ that satisfies three laws:
\begin{enumerate}
\item $\forall \ia{s, p}.\, t\, (s \dn q\, p) = t\, s \dnp p$,
\item $\forall \ia{s}.\, \o\, \ia{s} = q\, (\op\, \ia{t\, s})$, 
\item $\forall \ia{s, p, p'}.\, q\, p \pl \ia{s}\, q\, p' = q\, (p \plp \ia{t\, s}\, p')$.
\end {enumerate}
\end{defi}
In the special case $S = S' = 1$, laws 2 and 3 are the laws of a
monoid morphism.

Recall the intuition that $t$ determines the shape of the
data-structure that some given data-structure is sent to and $q$
identifies for every position in the data-structure returned a
position in the given data-structure. These laws say that the
positions in the sub-data-structure for any position in the resulting
data-structure must map back to positions in the corresponding
sub-data-structure of the given data-structure. This means that they
can receive data only from those positions, other flows are
forbidden. Morphisms between directed containers representing
node-labelled tree datatypes are exactly upwards accumulations---this
was one of the motivations for choosing the name `directed
containers'.

\begin{exa}
  The container representations of the head and drop-even functions
  for non-empty lists are directed container morphisms for the
  directed container of non-empty lists and suffixes (and the identity
  directed container). But those of self-append and reversal are not.
\end{exa}

\begin{exa}
  For the directed container of non-empty lists and cyclic shifts, not
  only the representations of the head and drop-even functions but
  also the self-append function are directed container morphisms.
\end{exa}

The identities and composition of $\Cont$ can give the identities and
composition for directed containers, since for every directed
container $E = (C, \dn, \o, \pl)$, the identity container morphism
$\cid\, \ia{C}$ is a directed container morphism and the composition
$h \ccomp h'$ of two directed container morphisms is also a directed
container morphism.

\begin{prop}
  Directed containers form a category $\DCont$.
\end{prop}


\subsection{Interpretation of Directed Containers}

As directed containers are containers with some operations obeying
some laws, a directed container should denote not just a set functor,
but a set functor with operations obeying some laws. The correct
domain of denotation for directed containers is provided by comonads
on sets.

\begin{defi}
Given a directed container $E = (S \lhd P, \dn, \o, \pl)$, we define
its \emph{interpretation} $\dcsem{E}$ to be the set functor $D =
\csem{S \lhd P}$ (i.e., the interpretation of the underlying container)
together with two natural transformations
\[
\begin{array}{l}
\eps : \forall \ia{X}.\, (\Sigma s: S.\, P\, s \to X) \to X\\
\eps\, (s, v) = v\, (\o\, \ia{s})\\
\de : \forall \ia{X}.\, (\Sigma s: S.\, P\, s \to X) \to \Sigma s: S.\, P\, s \to \Sigma s': S.\, P\, s'\to X\\
\de\, (s, v) = (s, \lambda p.\, (s \dn p, \lambda p'.\, v\, (p \pl \ia{s}\, p')))
\end{array}
\]
The directed container laws ensure that the natural transformations
$\eps$, $\de$ make the counit and comultiplication of a comonad
structure on $D$.
\end{defi}

Intuitively, the counit extracts the data at the root position of a
data-structure (e.g., the head of a non-empty list), the
comultiplication, which produces a data-structure of data-structures,
replaces the data at every position with the sub-data-structure
corresponding to this position (e.g., the corresponding suffix or
cyclic shift).

The interpretation $\dcsem{h}$ of a morphism $h$ between directed
containers \linebreak $E = (C, {\dn}, \o, {\pl})$, $E' = (C', \dnp,
\op, \plp)$ is defined by $\dcsem{h} = \csem{h}$ (using that $h$ is a
container morphism between $C$ and $C'$). The directed container
morphism laws ensure that this natural transformation between
$\csem{C}$ and $\csem{C'}$ is also a comonad morphism between
$\dcsem{E}$ and $\dcsem{E'}$.

Since the category $\Comonad{\SetB}$ inherits its identities and composition from \linebreak 
$[\SetB, \SetB]$, the functor $\dcsem{-}$ also preserves the identities and
composition.

\begin{prop}
$\dcsem{-}$ is a functor from $\DCont$ to $\Comonad{\SetB}$.
\label{prop:dcsemfunctor}
\end{prop}

Similarly to the case of natural transformations between container
interpretations, one can also ``quote'' comonad morphisms between
directed container interpretations into directed container
morphisms.\ For any directed containers $E = (C, {\dn}, \o, {\pl})$, $E' =
(C', {\dnp}, \op, {\plp})$ and any morphism $\tau$ between the comonads
$\dcsem{E}$ and $\dcsem{E'}$ (which is a natural transformation
between $\csem{C}$ and $\csem{C'}$), the container morphism
$\dcquo{\tau} = \cquo{\tau}$ between the underlying containers $C$ and
$C'$ is also a directed container morphism between $E$ and $E'$. The
directed container morphism laws follow from the comonad morphism
laws.

From what we already know about interpretation and quoting of
container morphisms, it is immediate that $\dcquo{\dcsem{h}} = h$ for
any directed container morphism $h$ and that $\dcquo{\tau} =
\dcquo{\tau'}$ implies $\tau = \tau'$ for any comonad morphisms
$\tau$ and $\tau'$ between directed container interpretations.

\begin{prop}
\label{prop:dcsemfullyfaithful}
$\dcsem{-}$ is fully faithful.
\end{prop}

The \emph{identity container} $\cId = 1 \lhd \lambda \zt.\, 1$ extends
trivially to an identity directed container whose denotation is
isomorphic to the identity comonad.  But, similarly to the situation
with functors and comonads, composition of containers fails to yield a
composition monoidal structure on $\DCont$.

We have elsewhere \cite{AU} shown that, similarly to the functors and
comonads case \cite{Beck}, the composition of the underlying
containers of two directed containers carries a \emph{compatible}
directed container structure if and only if there is a
\emph{distributive law} between these directed containers. Compatible
compositions of directed containers turn out to generalize
Zappa-Sz\'ep products of monoids~\cite{Zappa,Brin}, with distributive
laws playing the role of matching pairs of mutual actions.


\subsection{Containers \texorpdfstring{$\cap$}{n} Comonads
  \texorpdfstring{$=$}{=} Directed Containers} 
\label{sec:pullback}

Since not every functor can be represented by a container, there is no
point in asking whether every comonad can be represented as a directed
container. An example of a natural comonad that is not a directed
container is the cofree comonad on the finite powerset functor
$\mathcal{P}_\mathrm{f}$ (node-labelled nonwellfounded
strongly-extensional trees) where the carrier of this comonad is not a
container ($\mathcal{P}_\mathrm{f}$ is also not a container).  But,
what about those comonads whose underlying functor is an
interpretation of a container? It turns out that any such comonad does
indeed define a directed container that is obtained as follows.

Given a comonad $(D, \eps, \de)$ and a container $C = S \lhd P$ such
that $D = \csem{C}$, the counit $\eps$ and comultiplication $\de$
induce container morphisms
\[
\begin{array}{l}
\heps : C \to \cId\\
\heps = \teps \lhd \qeps = \cquo{\ee \comp \eps}\\
\hde :C \to C \cComp C \\
\hde = \tde \lhd \qde = \cquo{\mm\, \ia{C,C} \comp \de}
\end{array}
\]
using that $\csem{-}$ is fully faithful. From $(D, \eps, \de)$
satisfying the laws of a comonad we can prove that $(C, \heps, \hde)$
satisfies the laws of a comonoid in $\Cont$ (i.e., an object in $\mathbf{Comonoids}(\Cont)$). Further, we can define
\[
\begin{array}{l}
s \dn p = \snd\, (\tde\, s)\, p\\
\o\, \ia{s} = \qeps \ia{s}\, \zt\\
p \pl \ia{s}\, p' = \qde\, \ia{s}\, (p, p')
\end{array}
\]
and the comonoid laws further enforce the laws of the directed
container for $(C, {\dn}, \o, {\pl})$.  

It may seem that the maps $\teps$ and $\fst \comp \tde$ are not used
in the directed container structure, but $\teps : S \to 1$ contains no
information ($\forall \ia{s}.\, \teps\, s = \zt$) and the
comonad/comonoid right counital law forces that $\forall \ia{s}.\, \fst\,
(t^\de\, s) = s$, which gets used in the proofs of each of the five
directed container laws.  The latter fact is quite significant. It
tells us that the comultiplication $\de$ of any comonad whose
underlying functor is the interpretation of a container preserves the
shape of a given data-structure as the outer shape of the data-structure
returned.

The situation is summarized as follows.

\begin{prop}
\label{prop:comonad2dcontainer}
Any comonad $(D, \eps, \de)$ and container $C$ such that $D
= \csem{C}$ determine a directed container $\lceil (D, \eps,
\de), C \rceil$.
\end{prop}

\begin{prop}
\label{prop:pullbacklaw2}
$\lceil \dcsem{C,\dn,\o,\pl}, C \rceil = (C,\dn,\o,\pl)$.
\end{prop}

\begin{prop}
\label{prop:pullbacklaw1}
$\dcsem{\lceil (D, \eps, \de), C\rceil} = (D, \eps, \de)$.
\end{prop}

These observations combine into the following theorem.

\begin{prop}
\label{prop:pullback}
The following is a pullback in $\mathbf{CAT}$:
\[
\xymatrix@C=4em@R=4em{
\DCont  \ar[r]^{U} \ar[d]_{\dcsem{-}}^{\mathrm{f.f.}}
  & \Cont \ar[d]_{\csem{-}}^{\mathrm{f.f.}} \\
\Comonad{\SetB} \ar[r]^{U}
  & \SFun
}
\]
\end{prop}

\noindent A structured way to prove this theorem is to first note that a
pullback is provided by $\mathbf{Comonoids}(\Cont)$ and then verify
that $\mathbf{Comonoids}(\Cont)$ is isomorphic to $\DCont$.

Sam Staton pointed it out to us that the proof of the first part only
hinges on $\Cont$ and $[\SetB,\SetB]$ being monoidal categories and
$\csem{-} : \Cont \to [\SetB,\SetB]$ being a fully faithful monoidal
functor. Thus we actually establish a more general fact, viz., that
for any two monoidal categories $\mathcal{C}$ and $\mathcal{D}$ and a
fully-faithful monoidal functor $F : \mathcal{C} \to \mathcal{D}$, the
pullback of $F$ along the forgetful functor $U :
\mathbf{Comonoids}(\mathcal{D}) \to \mathcal{D}$ is
$\mathbf{Comonoids}(\mathcal{C})$.

In summary, we have seen that the interpretation of a container
carries the structure of a comonad exactly when it extends to a
directed container.


\section{Constructions of Directed Containers}
\label{sec:constructions}

We now show some constructions of directed containers. While some
standard constructions of containers extend to directed containers,
others do not.

\subsection{Coproducts of Directed Containers}

Given two directed containers $E_0 = (S_0 \lhd P_0, \dn_0, \o_0,
\pl_0)$, $E_1= (S_1 \lhd P_1, \dn_1, \o_1, \pl_1)$, their coproduct is
$E = (S \lhd P, \dn, \o, \pl)$ where the underlying container $C = S
\lhd P$ is the coproduct of containers $C_0 = S_0 \lhd P_0$ and $C_1 =
S_1 \lhd P_1$.  All of the directed container operations are defined
either using $\dn_0, \o_0, \pl_0$ or $\dn_1, \o_1, \pl_1$ depending on
the given shape.  This means that the subshape operation is given by
$\inl\, s \dn p = \inl\, (s \dn_0 p)$ and $\inr\, s \dn p = \inr\, (s
\dn_1 p)$, the root position is given by $\o\, \ia{\inl\, s} = \o_0\,
\ia {s}$ and $\o\, \ia{\inr\, s} = \o_1\, \ia {s}$ and the subshape
position translation operation is given by $p \pl \ia{\inl\, s}\, p' =
p \pl_0 \ia{s}\, p'$ and $p \pl \ia{\inr\, s}\, p' = p \pl_1 \ia{s}\,
p'$. The interpretation of $E$ is isomorphic to the coproduct of
comonads $\dcsem{E_0}$ and $\dcsem{E_1}$.

\begin{prop}
  $E$ defined above is a coproduct of the given directed containers
  $E_0$ and $E_1$.  It interprets to a coproduct of the comonads
  $\dcsem{E_0}$ and $\dcsem{E_1}$, whose underlying functor is
  isomorphic to $\csem{C_0}+\csem{C_1}$.
\end{prop}


\subsection{Products of (Strict) Directed Containers}
\label{sec:products}

There is no general way to endow the product of the underlying
containers of two directed containers $E_0 = (S_0 \lhd P_0, \dn_0,
\o_0, \pl_0)$ and $E_1= (S_1 \lhd P_1, \dn_1, \o_1, \pl_1)$ with the
structure of a directed container. One can define $S = S_0 \times S_1$
and $P\, (s_0,s_1) = P_0\, s_0 + P_1\, s_1$, but there are two choices
$\o_0$ and $\o_1$ for $\o$. Moreover, there is no general way to define
$p \pl p'$. But this should not be surprising, as the product of the
underlying functors of two comonads is not generally a comonad.  Also,
the product of two comonads would not be a comonad structure on the
product of the underlying functors. 

However, for monads it is known that, although the coproduct of two
arbitrary monads may not always exist and is generally relatively
difficult to construct explicitly \cite{kelly}, there is a feasible
explicit formula for the coproduct of two ideal monads \cite{ideal}.
The duality with comonads gives a formula for the product of two
coideal comonads.  

\begin{defi}
  A \emph{coideal comonad} on $\SetB$ is given by a functor $D^+ : \SetB
  \to \SetB$ and a natural transformation $\depl : D^+ \to D^+ \Comp D$
  such that the diagrams below commute
\[
\xymatrix@C=3em@R=3em{
D^+   \ar@{=}[dr] \ar[r]^{\depl} & D^+ \Comp D \ar[d]^{D^+ \Comp \eps}
& D^+   \ar[d]_{\depl} \ar[r]^{\depl} & D^+ \Comp D \ar[d]^{D^+ \Comp \de} \\
& D^+
& D^+ \Comp D  \ar[r]_{\depl \Comp D} & D^+ \Comp D \Comp D
}
\]
for a functor $D : \SetB \to \SetB$ and natural transformations $\eps: D
\to \Id$ and $\de : D \to D \Comp D$ defined by 
\begin{itemize}
\item $D X = X \times D^+ X$
\item $\eps : \forall\ia{X} .\, X \times D^+ X \to X$\\
          $\eps = \fst$
\item $\de : \forall \ia{X} .\, X \times D^+ X \to D\,X \times D^+\, (D\, X)$\\
          $\de = \langle \id , \depl \comp \snd \rangle$
\end{itemize}
\end{defi}\smallskip

\noindent The design of this definition ensures that the data $(D, \eps, \de)$
make a comonad as soon as the data $(D^+, \depl)$ satisfy the coideal
comonad laws.\footnote{The term `coideal comonad' is motivated by
  $(D^+, \depl)$ being a right comodule of the comonad $(D, \eps, \de)$. For
  the same concept, also the term `ideal comonad' has been used.}

Given two coideal comonads $(D_0^+, \depla)$ and $(D_1^+, \deplb)$, the
functor $D$ given by
\begin{itemize}
\item $D\, X = \carrierplao\, X \times \carrierplbo\, X$
\end{itemize}
where
\begin{itemize}
\item $(\carrierplao\, X , \carrierplbo\, X) = \nu (Z_0 , Z_1) .\, (D_0^+\, (X \times Z_1) , D_1^+\, (X \times Z_0))$
\end{itemize}
(assuming the existence of the final coalgebra) carries a coideal
comonad structure that is a product, in the category of all comonads,
of the given ones.

Next we define the corresponding specialization
of directed containers and give an explicit product construction for
this case. A strict directed container is, intuitively, a directed container
where no position in a non-root subshape of a shape translates to its
root, i.e., $p \pl p'$ should not be $\o$ when $p \neq \o$. 

\begin{defi}
A \emph{strict directed container} is specifiable by the data
\begin{itemize}
\item $S : \Set$
\item $P^+ : S \to \Set$
\item $\dnpl : \Pi s : S .\, P^+\, s \to S$
\item $\plpl : \Pi \ia{s : S}.\, \Pi p : P^+\, s.\, P^+\, (s \dnpl p) \to P^+\, s$
\end{itemize}
satisfying the laws
\begin{enumerate}
\item $\forall \ia{s,p,p'} .\, s \dnpl (p \plpl p') = (s \dnpl p) \dnpl p'$
\item $\forall \ia{s,p,p',p''} .\, (p \plpl \ia{s}\, p') \plpl p'' = p \plpl (p' \plpl p'')$
\end{enumerate}
\end{defi}

\noindent It induces a directed container $(S \lhd P , \dn , \o , \pl)$ via
\begin{itemize}
\item $P\, s = \Maybe\,(P^+ s)$
\item $s \dn \nothing = s$ \\
  $s \dn \just\, p = s \dnpl p$
\item $\o = \nothing$
\item $\nothing \pl p = p$ \\
  $\just\, p \pl \nothing = \just\, p$ \\
  $\just\, p \pl \just\, p' = \just\, (p \plpl p')$
\end{itemize}
Similarly to coideal comonads, the design of this definition also ensures that 
the data \linebreak $(S \lhd P , {\dn} , \o , \pl)$ make a directed container as soon as 
the data $(S \lhd P^+ , \dnpl , \plpl)$ satisfy
the strict directed container laws.\footnote{You may notice a small ``mismatch''
  between the definitions of strict directed containers and coideal
  comonads. We have given $\plpl$ the type $\Pi \ia{s : S}.\, \Pi p:
  P^+\, s.\, P^+\, (s \dnpl p) \to P^+\, s$ while the $\depl$ has type
  $D^+ \Comp D \to D^+$, not $D^+ \Comp D^+ \to D^+$. The reason is
  that the first option for the type of $\depl$ is more general and
  really the ``correct'' one for comonads.  For comonads whose
  underlying functors are containers, however, the corresponding type
  $\Pi \ia{s : S}.\, \Pi p: P^+\, s.\, P\, (s \dnpl p) \to P^+\, s$
  buys no additional generality.}

  Strict directed containers are the pullback of the interpretation of
  directed containers and the inclusion of coideal comonads into
  comonads.

Notice that the special case $S = 1$ describes monoids without right-invertible
non-unit elements (such monoids are trivially also without left-invertible non-unit elements; they arise from adding a unit to a semigroup freely).  For example, the datatype of lists and suffixes is a strict directed
container; on the other hand, the datatype of lists and cyclic shifts is not.

We take inspiration from the construction of the product of two coideal
comonads and construct the product of two strict directed
containers. 

Given two strict directed containers $E_0 = (S_0 \lhd P_0^+, {\dnpla},
{\plpla})$ and $E_1 = (S_1 \lhd P_1^+, {\dnplb}, {\plplb})$, we define 
the data $E = (S \lhd P^+, {\dnpl}, {\plpl})$ by
\begin{itemize}
\item $S = \shapeao \times \shapebo$ \\[2pt]
where \\[6pt]
$(\shapeao, \shapebo) = \nu (Z_0, Z_1).\, (\Sigma s_0 : S_0 .\, P_0^+ s_0 \to Z_1, \Sigma s_1 : S_1 .\, P_1^+ s_1 \to Z_0)$\\[3pt]

\item $P^+ (s_0 , s_1) = \posplao s_0 + \posplbo s_1$ \\[2pt]
where \\[6pt]
$(\posplao, \posplbo) = \mu (Z_0, Z_1).\, \newline (\lambda (s_0 , v_0).\, \Sigma p_0 : P_0^+\, s_0.\, \Maybe\, (Z_1\, (v_0\, p_0)), \lambda (s_1 , v_1).\, \Sigma p_1 : P_1^+ s_1 .\, \Maybe\, (Z_0\, (v_1\, p_1)))$\\[3pt]

\item $\dntpl : \Pi s : S .\, P^+ s \to S$ \\
$(s_0, s_1) \dntpl \inl\, p = s_0 \dnpll p$ \\
$(s_0, s_1) \dntpl \inr\, p = s_1 \dnplr p$ \\[2pt]
where \\[6pt]
$\dnpll : \Pi s : \shapeao .\, \posplao\, s \to S$ \\
$\dnplr : \Pi s : \shapebo .\, \posplbo\, s \to S$ \\
(by mutuual recursion) \\
	 $(s_0 , v_0) \dnpll (p_0 , \nothing) = ((s_0 \dnpla p_0 , \lambda p .\, v_0\, (p_0 \plpla p)) , v_0\, p_0)$\\
	 $(s_0 , v_0) \dnpll (p_0 , \just\, p) = v_0\, p_0 \dnplr p$ \\
	 $(s_1 , v_1) \dnplr (p_1 , \nothing) = (v_1\, p_1 , (s_1 \dnplb p_1 , \lambda p .\, v_1\, (p_1 \plplb p)))$\\
	 $(s_1 , v_1) \dnplr (p_1 , \just\, p) = v_1\, p_1 \dnpll p$\\[6pt]

\item $\pltpl : \Pi \ia{s : S} .\, \Pi p : P^+ s . P^+ (s \dnpl p) \to P^+ s$\\
          $\inl\, p \pltpl p' = \inl\, (p \plpll p')$\\
          $\inr\, p \pltpl p' = \inr\, (p \plplr p')$ \\[2pt]
where \\[6pt]
$\plpll : \Pi \ia{s : \shapeao} .\, \Pi p : \posplao\, s.\, P^+ (s \dnpll  p) \to \posplao\, s$ \\
$\plplr : \Pi \ia{s : \shapebo} .\, \Pi p : \posplbo\, s .\, P^+ (s \dnplr p) \to \posplbo\, s$ \\
(by mutual recursion) \\
          $(p_0 , \nothing) \plpll \inl\, (p_0' , p_1') = (p_0 \plpla p_0' , p_1')$\\
          $(p_0 , \nothing) \plpll \inr\, p  = (p_0 , \just\, p)$\\
          $(p_0 , \just\, p_1) \plpll p = (p_0 , \just\, (p_1 \plplr p))$ \\
          $(p_1 , \nothing) \plplr \inr\, (p_1' , p_0') = (p_1 \plplb p_1' , p_0')$\\
          $(p_1 , \nothing) \plplr \inl\, p  = (p_1 , \just\, p)$\\
          $(p_1 , \just\, p_0) \plplr p = (p_1 , \just\, (p_0 \plpll p))$ 
\end{itemize}

\begin{prop} \label{prop:prod} $E$ is a product, in the
  category of all directed containers, of the strict directed
  containers $E_0$ and $E_1$. It interprets to a product, in the
  category of all comonads, of their interpreting coideal comonads.
\end{prop}

The definitions above a considerable amount of detail, but the
intuition behind them is not difficult. The product of two strict
directed containers generalizes the coproduct of two monoids without
non-unit right-invertible elements. The elements of this monoid are
finite alternating sequences of non-unit elements of the two given
monoids.  The definitions above arrange for alternations of a similar
nature.


\subsection{Cofree Directed Containers}  
\label{sec:cofree}

Given a container $C_0 = S_0 \lhd P_0$, let us define $E = (S \lhd P,
\dn, \o, \pl)$ by
\begin{itemize}
\item $S = \nu Z.\, \Sigma s: S_0. P_0\, s \to Z$ 
\item $P = \mu Z.\,\lambda (s, v).\,1 + \Sigma p: P_0\, s.\, Z\, (v\, p)$ 
\item (by recursion) \\
      $(s, v) \dn \inl\, \zt = (s, v)$ \\
      $(s, v) \dn \inr\,(p, p')= v\, p \dn p'$
\item $\o\, \ia{s, v} = \inl\, \zt$ 
\item (by recursion) \\
      $\inl\, \zt \pl \ia{s, v}\,p'' = p''$ \\
      $\inr\, (p, p') \pl \ia{s, v}\, p'' = \inr\, (p, p' \pl \ia{v\, p}\, p'')$
\end{itemize}

\begin{prop} \label{prop:cofree} $E$ is a cofree directed container on
  $C_0$. It interprets into a cofree comonad on the functor $\csem{C_0}$,
  which has its underlying functor isomorphic to $D\, X = \nu Z.\, X
  \times \csem{C_0}\, Z$.
\end{prop}

\begin{exa}
In the special case $S_0 = 1$, we get that $S = \nu Z.\ \Sigma \zt :
1.\, P_0\, \zt \to Z \cong 1$ and this example degenerates to the free
monoid on a given set $P_0\, \zt$, i.e., the monoid of lists over
$P_0\, \zt$ (with the empty list as the unit and concatenation as the
multiplication operation). This directed container interprets into the
comonad of nonwellfounded node-labelled $P_0\, \zt$-branching trees.
\end{exa}


\subsection{Cofree Recursive Directed Containers} 

A recursive comonad is a coideal comonad $(D^+, \de^+)$ such that, for
any map $f : D^+\, (X \times Y) \to Y$, there exists a unique map
$f^\dagger : D^+\, X \to Y$ such that
\[
\xymatrix@C=6pc{
D^+\, X \ar[r]^{f^\dagger} \ar[d]_{\de^+}
  & Y \\
D^+\, (D\, X) \ar[r]_{D^+\, (X \times f^\dagger)}
  & D^+\, (X \times Y) \ar[u]_{f}
}
\]

Recursive directed containers are the pullback of the interpretation
of strict directed containers and the inclusion of recursive comonads
into coideal comonads.

Now the cofree recursive directed container on a given container $C$
is obtained by replacing the $\nu$ in the definition of the shape set
$S$ of the cofree directed container with $\mu$. The interpretation
has its underlying functor isomorphic to $D\, X = \mu Z.\, X \times
\csem{C}\, Z$, which is the cofree recursive comonad on $\csem{C}$.

While cofree directed containers represent datatypes of node-labelled
nonwellfounded trees, cofree recursive directed containers correspond
to node-labelled wellfounded trees. The simplest interesting example
is the datatype of non-empty lists (with its suffixes structure),
which is represented by the cofree recursive directed container on the
``maybe'' container $1+1 \lhd \lambda \{(\inl\, \zt) .\, 0 ~;~ (\inr\, \zt) .\, 1\}$, 
i.e., two shapes, one with no positions, the other with one position.


\subsection{Data-structures with a Focus}

Below we discuss directed containers equipped a 
notion of focus. We present a construction for turning any container into 
a directed container with a designated focus. We also show that 
the zipper types of  Huet~\cite{Huet1997} have a direct
representation as directed containers.

\subsubsection*{Focussing}

Any container $C_0 = S_0 \lhd P_0$ defines a directed container $E = ({S
  \lhd P}, {\dn}, \o, {\pl})$ as follows. We take $S = \Sigma s :
S_0.\, P_0\, s$, so that a shape is a pair of a shape $s$, the ``shape
proper'', and an arbitrary position $p$ in that shape, the ``focus''.
We take $P\, (s, p) = P_0\, s$, so that a position in the shape $(s,
p)$ is a position in the shape proper $s$, irrespective of the focus.
The subshape determined by position $p'$ in shape $(s, p)$ is given by
keeping the shape proper but changing the focus: $(s, p) \dn p' = (s,
p')$. The root in the shape $(s, p)$ is the focus $p$, so $\o\,
\ia{s, p} = p$.  Finally, we take the translation of positions from
the subshape $(s, p')$ given by position $p'$ to shape $(s, p)$ to be
the identity, by defining $p' \pl \ia{s, p}\, p'' = p''$.  All
directed container laws are satisfied.  

The directed container $E$ so obtained interprets into the canonical
comonad structure on the functor $\partial \csem{C_0} \times \Id$, where
$\partial F$ denotes the derivative of the functor $F$. (For
derivatives of set functors and containers, see Abbott et
al.~\cite{abbott.altenkirch.ghani.mcbride:data}.)

Differently from, e.g., the cofree directed container construction,
this construction is not a functor from $\Cont$ to $\DCont$. Instead,
it is a functor from the category of containers and Cartesian
container morphisms (where position maps are bijections).

\subsubsection*{Zippers}
Inductive (tree-like) datatypes with a designated focus position are
isomorphic to the zipper types of Huet~\cite{Huet1997}. A zipper
data-structure encodes a tree with a focus as a pair of a context and a
tree. The tree is the subtree of the global tree rooted by the focus
and the context encodes the rest of the global tree. On zippers,
changing the focus is supported via local navigation operations for
moving one step down into the tree or up or aside into the context.

Zipper datatypes are directly representable as directed containers. We
illustrate this on the example of zippers for lists (which
are, in fact, the same as zippers for non-empty lists, as one cannot focus
on a position in the empty list). 

\begin{exa}
A list zipper is a pair of a list
(the context) and a non-empty list (the suffix determined by the focus
position).  Accordingly, by defining $S = \Nat \times \Nat$, the shape
of a zipper is a pair $(s_0, s_1)$ where $s_0$ is the shape of the
context and $s_1$ is the shape of the suffix. For positions, it is
convenient to choose $P\, (s_0, s_1) = \{-s_0,\ldots, s_1\}$ by
allocating the negative numbers in the interval for positions in the
context and non-negative numbers for positions in the suffix.  The
root position is $\o\, \ia{s_0, s_1} = 0$, i.e., the focus. The
subshape for each position is given by $(s_0, s_1) \dn p = (s_0 + p,
s_1 - p)$ and translation of subshape positions by $p \pl \ia{s_0,
  s_1}\, p' = p + p'$.

Fig.~\ref{fig:dcontainernelistfocus} gives an example of a non-empty
list with focus with its shape fixed to $s = (5,6)$. It should be
clear from the figure how the $\pl$ operation works on positions $p =
4$ and $p' = -7$ to get back the position $p \pl p' = -3$ in the
initial shape. The subshape operation $\dn$ works as follows: $s \dn
p$ gives back a subshape $s' = (9,2)$ and $s \dn (p \pl p')$ gives
$s'' = (2,9)$.

\begin{figure}[h]
\small
\[
\xymatrix@C=1.55em@R=1.2em{
& & & & &  &  & &  &  &  &  &  &\\
& & & & &  &  & &  &  &  &  &  &\\
\bullet & \bullet & \bullet & \bullet  & \bullet  & & \bullet  \ar@{->}@/_1pc/[rrrrr]_{p = 4} \ar@{->}@/^1pc/[llll]^>>>>>>>{p \pl p =  + p' = -3} \ar@{-}[uurr]_>{s = (5,6)}  \ar@{-}[ddrr] \ar@{-}[uull] \ar@{-}[ddll] & & \bullet  & \bullet  & \bullet  & \bullet  \ar@{.>}@/_1pc/[lllllllll]_>>>>>>>>{p' = -7}  \ar@{.}[uurr]_>{s' = (9,2)} \ar@{.}[ddrr] \ar@{.}[uull] \ar@{.}[ddll] & \bullet & \bullet\\
& & & & &  &  & &  &  &  &  &  &\\
& & & & &  &  & &  &  &  &  & &
}
\]
\caption{The shape and positions for non-empty lists of length 12 focussed at position 5}
\label{fig:dcontainernelistfocus}
\end{figure}

The isomorphism of the directed container representation of the
list zipper datatype and the directed container of focussed lists is
$t : \Nat \times \Nat \to \Sigma s : \Nat.\, \{0, \ldots, s-1\}$, $t\,
(s_0, s_1) = (s_0 + s_1 + 1, s_0)$, $q : \Pi \ia{(s_0, s_1) : \Nat
  \times \Nat}.\, \{0, \ldots, s_0 + s_1\} \to \{-s_0, \ldots,
s_1\}$, $q\, \ia{s_0, s_1}\, p = p - s_0$.
\end{exa}

We refrain here from delving deeper into the topic of derivatives and
zippers, leaving this discussion for another occasion.


\section{Containers \texorpdfstring{$\cap$}{n} Monads
  \texorpdfstring{$=$}{=} ?}
\label{sec:monads}

Given that comonads whose underlying functor is the interpretation of
a container are the same as directed containers, it is natural to ask
whether a similar characterization is possible for monads whose
underlying functor can be represented as a container. The answer is
``yes'', but the additional structure is more involved than that of
directed containers.

Given a container $C = S \lhd P$, the structure $(\eta, \mu)$ of a
monad on the functor $T = \csem{C}$ is interdefinable with the
following structure on $C$
\begin{itemize*}
\item $\mathsf{e} : S$ (for the shape map for $\eta$), 
\item $\bullet : \Pi s : S. (P\, s \to S) \to S$ (for the shape map for $\mu$),
\item ${\nnwarrow} : \Pi \ia{s : S}.\, \Pi v : P\, s \to S. P\, (s \bullet v) \to P\, s$ and  
\item ${\nnearrow} : \Pi \ia{s : S}.\, \Pi v : P\, s \to S. \Pi p: P\, (s \bullet v).\,  P\, (v\, (v \nnwarrow \ia{s} p))$ (both for the position map for $\mu$)
\end{itemize*}
subject to three shape equations and five position equations. 
Perhaps not unexpectedly, this amounts to having a monoid structure on $C$.
We refrain from a more detailed discussion of this variation of the
concept of containers.

\begin{exa}
To get some intuition, consider the monad structure on the datatype of
lists.  The unit is given by singleton lists and multiplication is
flattening a list of lists by concatenation. For the list container $S
= \Nat$, $P\, s = \Fin\, s$, we get that $\mathsf{e} = 1$, $s \bullet
v = \sum_{p : \Fin\, s} v\, p$, $v\, \nnwarrow \ia{s}\, p =
\mathrm{[greatest~} p': \Fin\, s \mathrm{~such~that~} \sum_{p'' :
  \Fin\, p'} v\, p'' \leq p \mathrm{]}$ and $v\, \nnearrow \ia{s}\,
p = p - \sum_{p'' : \Fin\, (v \nnwarrow \ia{s}\, p)} v\, p''$. The
reason is that the shape of singleton lists is $\mathsf{e}$ while
flattening a list of lists with outer shape $s$ and inner shape $v\,
p$ for every position $p$ in $s$ results in a list of shape $s \bullet
v$. For a position $p$ in the shape of the flattened list, the
corresponding positions in the outer and inner shapes of the given
list of lists are $v\, \nnwarrow \ia{s}\, p$ and $v \nnearrow \ia{s}\,
p$.
\end{exa}


\section{Cointerpreting Directed Containers into Monads}
\label{sec:cointerp}

What we have just described is not the only way to relate containers
to monads.  In a recent work \cite{AU:updmon}, we defined
\emph{cointerpretation} of containers as the functor $\ccosem{-} :
\Cont^\opc \to [\SetB, \SetB]$ given by
\[
\ccosem{S \lhd P} X = \Pi s : S.\, P s \times  X \cong (\Pi s : S.\, P\, s) 
\times (S \to X)
\]
Differently from $\csem{-}$, the functor $\ccosem{-}$ is neither full
nor faithful. It also fails to be monoidal for the monoidal structure
on $\Cont^\opc$ (taken from $\Cont$). But it is lax monoidal.

It is straightforward that $\DCont^\opc \cong (\Comonoid{\Cont})^\opc
\cong \Monoid{\Cont^\opc}$. Lax monoidal functors send monoids to
monoids. Hence $\ccosem{-}$ lifts to a functor $\dccosem{-} :
\DCont^\opc \to \Monad{\SetB}$ that equips each set functor $\ccosem{S
  \lhd P}$ with a monad structure
\[
\begin{array}{l}
\eta : \forall \ia{X}.\, X \to \Pi s : S.\, P\, s \times X \\
\eta\, x\, s = (\o\, \ia{s}, x) \\
\mu : \forall \ia{X}.\, 
   (\Pi s : S.\ P\, s \times \Pi s' : S.\ P\, s' \times  X) 
        \to \Pi s : S.\, P\, s \times X \\
\mu\, f\, s = \mathsf{let~} \{(p, g) = f\, s;\, (p', x) = g\, (s \dn p)\}
                \mathsf{~in~} (p \pl p', x)
\end{array}
\]
Due to the resemblance to compatible compositions of reader and writer
monads, we call monads in the image of this functor ``dependently
typed update monads''. It is instructive to think of shapes in $S$ as
states, positions in $P s$ as updates applicable to a state $s$ (or
programs safe to evaluate from state $s$), $s \dn p$ as the result of
applying an update $p$ to the state $s$ (or the result of evaluating
$p$ from $s$), $\o\, \ia{s}$ as the nil update in state $s$ and $p \pl
p'$ as accumulation of two consecutive updates (skip and sequential
composition).

\begin{exa}
  The directed container for the nonempty list comonad, $S = \Nat$,
  $P\, s = [0..s]$, $s \dn p = s - p$, $\o = 0$, $p \pl p' = p + p'$,
  gives us a monad on the set functor $T$ given by $T\, X = \Pi s:
  \Nat.\, [0..s] \times X$. The states are natural numbers; the
  updates applicable to a state $s$ are numbers not greater than $s$;
  applying an update means decrementing the state.
\end{exa}

We can see that directed containers are not more ``comonadic''
inherently than they are ``monadic''.  We see them first of all as an
algebraic-like structure in their own right, a generalization of
monoids.


\section{Directed Containers in Categories with Pullbacks}
\label{sec:polycom}

Container theory can be carried out in locally Cartesian closed
categories (LCCCs)---the LCCC generalization of containers being well
known under the name of polynomials
\cite{gambino.hyland:poly,kock:polyfuncandtrees}---and even more
generally in categories with pullbacks \cite{weber}. It is natural to
expect the same of directed container theory.

This is the case indeed. The proofs in this paper can be seen as having been
carried out in the internal language of an LCCC (with the assumptions
of existence of initial algebras and final coalgebras corresponding to
assumptions about availability of W- and M-types).

In the weaker setting of a category with pullbacks, one has to be a
lot more careful. It is possible to define the concepts required from
the first principles. 

We show the definitions of the counterparts of directed containers and
directed container morphisms; we call them ``directed polynomials''
and ``directed polynomial morphisms'' in the local scope of this
section.

In all diagrams below, bullet-labelled nodes with a pair of unlabelled
outgoing arcs denote pullbacks defined by a pair of maps that are
given directly or constructed. Dashed arrows denote unique maps into a
pullback. The polygon actually required to commute is marked with a
small circular arrow.

Given a category with pullbacks $\mathcal{C}$, a directed polynomial is
given by
\begin{itemize}
\item two objects $S$ and $P$ (``sets'' of shapes and positions) and
  an exponentiable map $\sh$ (assigning every position a shape)
\[
\xymatrix@C=3em@R=3em{
P \ar[d]^{\sh}
\\
S
}
\]
\item a morphism $\dn$ picking out a shape for each position (the
  corresponding subshape)
\[
\xymatrix@C=3em@R=3em{
P \ar[r]_{\dn} & S
}
\]
\item a map $\o$ picking out, for every shape, a position in that
  shape (the root position)
\[
\xymatrix@C=3em@R=3em{
P \ar[d]_\sh &
\\
S \ar@{=}[r] \ar@{}[ur]|<<<<{\circlearrowright} & S \ar[ul]_{\o}
}
\]
\item a map $\pl$ sending a position in a given, global shape and a
  position in the corresponding to subshape to a position in the
  global shape (translation of the subshape position to the global
  shape)
\[
\xymatrix@C=3em@R=3em{
P \ar@/_1.5pc/[dddr]_{\sh}
\\ 
& \bullet \ar[ul]_{\pl} \ar[d] \ar[r]    
   \ar@{}[dl]|(0.4){\circlearrowleft}
       & P \ar[d]^{\sh}
\\
 & P \ar[r]_{\dn} \ar[d]^{\sh} & S
\\
& S
}
\]
\end{itemize}
satisfying the following five laws:
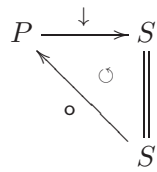
\begin{enumerate}
\item 
\[
\xymatrix@C=3em@R=3em{
P \ar[r]^{\dn} & S \ar@{=}[d] \ar@{}[dl]|<<<<<{\circlearrowleft} 
\\
       & S \ar[ul]^{\o}
}
\]
\item 
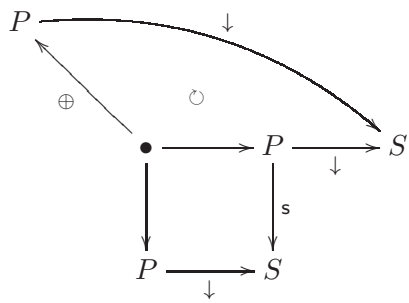
\[
\xymatrix@C=3em@R=3em{
P \ar@/^1.5pc/[drrr]^{\dn}
  & 
    &
\\ 
& \bullet \ar@{}[ur]|(0.4){\circlearrowright} \ar[ul]^{\pl} \ar[d] \ar[r]  
  & P \ar[d]^{\sh} \ar[r]_{\dn} 
    & S
\\
  & P \ar[r]_{\dn} 
   & S
}
\]
\item 
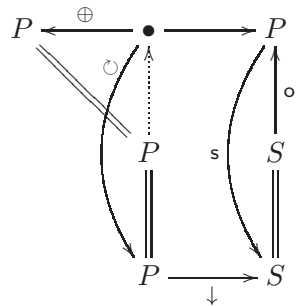
\[
\xymatrix@C=3em@R=3em{
P \ar@{=}[dr] & \bullet \ar@/_1.5pc/[dd] \ar[r]  \ar[l]_{\pl} \ar@{}[dl]|(0.3){\circlearrowright}  & P \ar@/_1.5pc/[dd]_{\sh} 
\\
& P \ar@{.>}[u] & S \ar[u]_{\o}
\\
& P \ar[r]_{\dn} \ar@{=}[u] & S \ar@{=}[u]
}
\]
\item 
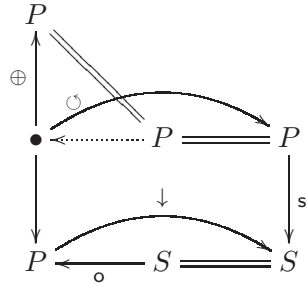
\[
\xymatrix@C=3em@R=3em{
P \ar@{=}[dr] &
\\
\bullet \ar[d] \ar@/^1.5pc/[rr] \ar[u]^{\pl} \ar@{}[ur]|(0.3){\circlearrowleft} & P \ar@{.>}[l] & P \ar[d]^{\sh} \ar@{=}[l]
\\
P \ar@/^1.5pc/[rr]^{\dn} & S \ar[l]^{\o} \ar@{=}[r] & S
}
\]
\item 
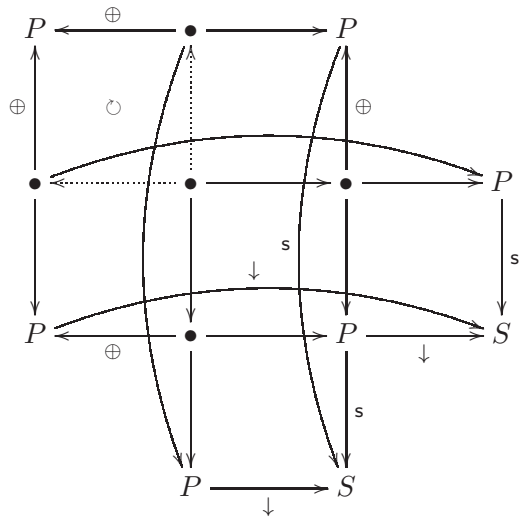
\[
\xymatrix@C=4em@R=4em{
P \ar@{}[dr]|-{\circlearrowright}  
& \bullet \ar[l]_{\pl}  \ar@/_1.5pc/[ddd] \ar[r] 
  & P \ar@/_1.5pc/[ddd]_(0.47){\sh}
\\
\bullet \ar[d] \ar@/^1.5pc/[rrr] \ar[u]^{\pl} 
& \bullet \ar[d]  \ar[r] \ar@{.>}[u] \ar@{.>}[l] 
  & \bullet \ar[d] \ar[r] \ar[u]_{\pl} 
    & P \ar[d]^{\sh} 
\\
P \ar@/^1.5pc/[rrr]^(0.47){\dn} 
& \bullet \ar[l]^{\pl} \ar[d] \ar[r] 
  & P \ar[r]_{\dn} \ar[d]^{\sh} & S
\\
& P \ar[r]_{\dn} 
  & S
}
\]
\end{enumerate}\medskip 

\noindent The data $S$, $\dn$, $\o$, $\pl$ here correspond to the homonymous data of
a directed container while $P$ and $\sh$ together correspond to $P$. The
five laws governing them correspond exactly to the five laws of a
directed container.

A \emph{morphism} between two directed polynomials $(S, P, \sh, \dn, \o,
\pl)$ and $(S', P', \sh', \dnp, \op, \plp)$ is given by two maps $t$ and
$q$ (of shapes and positions)
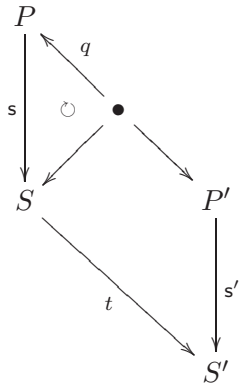
\[
\xymatrix@R=2em@C=2em{
P \ar[dd]_{\sh}
  & 
    & \\
\ar@{}[r]|-{\circlearrowright}
  & \bullet \ar[ul]_{q} \ar[dl] \ar[dr]
    & \\
S \ar[ddrr]_{t}
  & 
    & P' \ar[dd]^{\sh'} \\
  & 
    & \\
  & 
    & S'
}
\]
satisfying the following three laws:
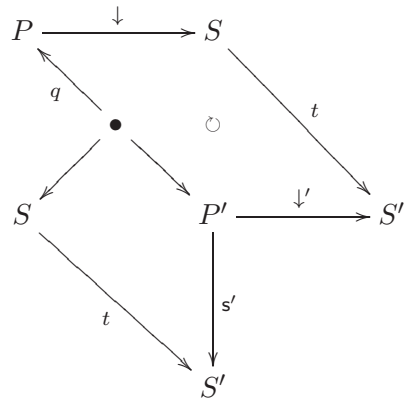
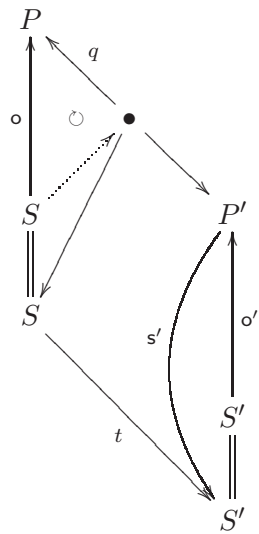
\begin{enumerate}
\item 
\[
\xymatrix@R=2em@C=2em{
P 
    \ar[rr]^{\dn}
  & 
    & S \ar[ddrr]^{t}
      &
        &\\
  & \bullet \ar[ul]^{q} \ar[dl] \ar[dr] \ar@{}[rr]|-{\circlearrowright}
    & 
      & 
        &\\
S \ar[ddrr]_{t}
  & 
    & P' \ar[dd]^{\sh'} \ar[rr]^{\dnp} 
      &
        & S' \\
  & 
    & 
      & 
        &\\
  & 
    & S' 
      &
        &
}
\]
\item 
\[
\xymatrix{
P \\
\ar@{}[r]|-{\circlearrowright}
  & \bullet \ar[ul]_{q} \ar[ddl] \ar[dr]  \\
S \ar[uu]^{\o} \ar@{=}[d] \ar@{.>}[ur]
  & & P' \ar@/_2pc/[ddd]_(0.4){\sh'}\\
S \ar[ddrr]_{t} \\
  & & S' \ar[uu]_{\op} \ar@{=}[d]\\
  & & S' 
}
\]
\item 
\[
\xymatrix@R=2em@C=2em{
& & P \\
\\
& & \ar@{}[rrrd]|(0.4){\circlearrowright}
    & & & \\
& & \bullet \ar[uuu]^{\pl} \ar[rr] \ar[dd]
    & & P \ar[dd]^{\sh}
        & \bullet \ar[uuulll]_(0.35){q} \ar@/^2.3pc/[ddddlll] \ar[dddrrr]\\
\\
& & P \ar[rr]_{\dn} \ar[dd]_{\sh}
    & & S \ar@/^1pc/[ddddddrrrrrr]^(0.4){t}\\
& & & & & \bullet \ar[dlll] \ar[dddrrr] \ar@{.>}[rr] \ar@{.>}[dd]
                   \ar@/^1pc/@{.>}[uuulll] \ar@{.>}[uuu]
          & & \bullet \ar[uuulll]_(0.35){q} \ar[ulll] \ar[dddrrr]
              & P' \ar@/_2pc/[ddddddd]_(0.45){\sh'} \\
& & S \ar[ddddddrrrrrr]_{t}
    & & & & & \\
& & & & & \bullet \ar[uuulll]^(0.35){q} \ar[ulll] \ar[dddrrr]
           \\
& & & & & & & & \bullet \ar[uuu]_(0.4){\plp} \ar[rr] \ar[dd]
                & & P' \ar[dd]^{\sh'}\\
\\
& & & & & & & & P' \ar[rr]_{\dnp} \ar[dd]^{\sh'}
                & & S' \\
\\
& & & & & & & & S'
}
\]
\end{enumerate}

The data $t$, $q$ correspond to the homonymous data of a directed
container morphism and the three laws to the three laws of a directed
container morphism.

In the special case $\mathcal{C} = \SetB$, the definitions of a
directed polynomial and directed polynomial morphism are equivalent to
those of a directed container and directed container morphism.

Remarkably, the definition of a directed polynomial is completely
symmetric in $\sh$ and $\dn$---swapping them over we also get a
directed polynomial. The definition of a directed polynomial morphism
is symmetric, if $q$ is an isomorphism.

The definitions of the interpretation of a directed polynomial
resp. directed polynomial morphism into a comonad resp. comonad
morphism require using distributivity pullbacks in $\mathcal{C}$ (or
pullbacks in its slice categories).


\section{Related Work}
\label{sec:related}

The core of this paper builds on the theory of containers as developed
by Abbott, Altenkirch and Ghani \cite{Abbott2005,abbott:phd} to
analyze strictly positive datatypes. Some generalizations of the
concept of containers are the indexed containers of Altenkirch and
Morris \cite{altenkirch.morris:indexed,morris:thesis} and the quotient
containers of Abbott et
al.~\cite{abbott.altenkirch.ghani.mcbride:quotient}. In our work we
look at a specialization of containers rather than a
generalization. Recently \cite{AU}, we have also studied compatible
compositions of directed containers and how they generalize 
Zappa-Sz\'ep products \cite{Zappa} of two monoids.

Simple/indexed containers are intimately related to strongly
positive datatypes/families and simple/dependent polynomial functors
as appearing in the works of Dybjer \cite{dybjer1997}, Moerdijk and
Palmgren \cite{moerdijk.palmgren}, Gambino and Hyland
\cite{gambino.hyland:poly}, Kock \cite{kock:polyfuncandtrees}.  Girard's normal functors
\cite{girard:normal} and Joyal's analytic functors
\cite{joyal:analytic} functors are similar to containers resp.\
quotient containers, but only allow for finitely many positions in a
shape. Gambino and Kock \cite{gambino.kock:monads} also treat
polynomial monads.

Abbott, Altenkirch, Ghani and McBride
\cite{abbott.altenkirch.ghani.mcbride:data} have investigated
derivatives of datatypes. Derivatives provide a systematic way to explain
Huet's zipper type \cite{Huet1997}.

Brookes and Geva \cite{bg92} and later Uustalu with coauthors
\cite{essence,atteval,tt,cell} have used comonads to analyze notions
of context-dependent computation such as dataflow computation,
attribute grammars, tree transduction and cellular automata. Uustalu
and Vene's \cite{uustalu.vene:comonadic} observation of a connection
between bottom-up tree relabellings and containers with extra
structure started our investigation into directed containers.


\section{Conclusions and Future Work}
\label{sec:concl}

We introduced directed containers as a specialization of containers
for describing a certain class of datatypes (data-structures where
every position determines a sub-data-structure) that occur very
naturally in programming. It was a pleasant discovery for us that
directed containers are an entirely natural concept also from the
mathematical point of view: they are the same as containers whose
interpretation carries the structure of a comonad. They also
generalize monoids in an interesting way. In a recent piece of work
\cite{AU:updlenses}, we have witnessed that coalgebras of comonads
interpreting directed containers are relevant for bidirectional
transformations as a flavor of lenses (``dependently typed update
lenses'').

As future work, we intend to take a closer look at focussing and
related concepts, such as derivatives. A curious special case of
directed containers supports translation of the root of a shape into
every subshape. Such bidirectional containers include, e.g., focussed
containers and generalize groups in the same way as directed
containers generalize monoids. We would like to find out if this
specialization of directed containers is an interesting and useful
concept. We wonder whether our explicit formula for the product of two
directed containers can be scaled to the general, non-strict, case.
Last, we would like to analyze containers that are monads more
closely.


\subsubsection*{Acknowledgments} 

We are indebted to Thorsten Altenkirch, Jeremy Gibbons, Peter Morris,
and Sam Staton for comments and suggestions. We thank our anonymous
referees for the useful feedback that helped us improve the article.


\appendix

\section{Proofs for Section~\ref{sec:dcontainers}}

\subsection*{Proof of Proposition \ref{prop:dcsemfunctor}}

\proof
We must check that the interpretation $\dcsem{E} = (D, \eps,
\de)$ of the given directed container $E = (C, \dn, \o, \pl)$ is a comonad.

\noindent Proof of the right counital law:
\[
\begin{array}{cl}
&D\, \eps\, (\de\,(s,v)) \\
= & \quad\{\textrm{definition of $D$}\} \\
&(\lambda (s,v).\, (s,\lambda  p.\, \eps\,(v\, p)))\, (\de\,(s,v))\\
=& \quad\{\textrm{definitions of $\eps$, $\de$}\}\\
&(s,\lambda p.\, v\,(p \pl \, \o))\\
=& \quad\{\textrm{directed container law 3}\}\\
&(s,v)
\end{array}
\]
\noindent Proof of the left counital law:
\[
\begin{array}{cl}
&\eps\, (\de\, (s,v))\\
=& \quad\{\textrm{definitions of $\eps$, $\de$}\}\\
&(s \dn \o, \lambda p'.\, v\,(\o \pl \, p'))\\
=& \quad\{\textrm{directed containers laws 1 and 4}\}\\
&(s , v)
\end{array}
\]
\noindent Proof of the coassociativity law:
\[
\begin{array}{cl}
& D\, \de\, (\de\,(s,v))\\
=& \quad\{\textrm{definition of $D$}\}\\
&(\lambda  (s,v).\, (s,\lambda  p.\, \de\,(v\, p)))\, (\de\,(s,v))\\
=& \quad\{\textrm{definition of $\de$}\}\\
&(s,\lambda  p.\, (s \dn p,\lambda p'.\, ((s \dn p) \dn p', \lambda p''.\, v\,(p \pl \, (p' \pl p'')))))\\
=& \quad\{\textrm{directed container laws 2 and 5}\}\\
&(s,\lambda  p.\, (s \dn p,\lambda p'.\, (s \dn (p \pl p'),\lambda p''.\, v\, ((p \pl p') \pl p''))))\\
= & \quad\{\textrm{definition of $\de$}\}\\
&\de\,(\de\,(s,v))
\end{array}
\]

We must also verify that the interpretation $\dcsem{h} = \tau$ of a
morphism $h = t \lhd q$ between two directed containers $E = (C, \dn,
\o, \pl)$ and $E' = (C', \dn', \o', \pl')$ is a comonad morphism
between $\dcsem{E} = (D,\eps,\de)$ and $\dcsem{E'} = (D',\eps',\de')$.

\noindent Proof of the counit preservation law:
\[
\begin{array}{cl}
&\eps\,(s,v) \\
=& \quad\{\textrm{definition of $\eps$}\}\\
&v\, \o\\
=& \quad\{\textrm{directed container morphism law 2}\}\\
&v\,(q\, \o')\\
=& \quad\{\textrm{definitions of $\tau$, $\eps'$}\}\\
&\eps'\,(\tau\, (s,v))
\end{array}
\]
\noindent Proof of the comultiplication preservation law:
\[
\begin{array}{cl}
&D\, \tau\, ( \tau\, (\de\,(s,v)))\\
=& \quad\{\textrm{definition of $D$}\}\\
&(\lambda (s,v).\, (s,\lambda p.\,  \tau\, (v\, p)))\, ( \tau\, (\de\,(s,v)))\\
=& \quad\{\textrm{definitions of $\tau$, $\delta$}\}\\
&(t\, s,\lambda p.\, (t\, (s \dn q\, p), \lambda p'.\, v\, (q\, p \pl  q\, p')))\\
=& \quad\{\textrm{directed container morphism laws 1 and 3}\}\\
&(t\, s,\lambda p.\, (t\, s \dn' p, \lambda p'.\, v\, (q\, (p \pl' p'))))\\
=& \quad\{\textrm{definitions of $\tau$, $\delta'$}\}\\
&\de'\, (\tau\, (s,v))\rlap{\hbox to 268 pt{\hfill\qEd}}
\end{array}
\] 

\subsection*{Proof of Proposition \ref{prop:dcsemfullyfaithful}}

\proof
From Proposition~\ref{prop:csemfullyfaithful}, we know that the interpretation 
of containers is fully faithful. It remains to show that, 
for directed containers $E = (C, \dn, \o,
\pl)$, $E' = (C', \dnp, \op, \plp)$ and a morphism $\tau$ between
the comonads $\dcsem{E}$ and $\dcsem{E'}$, the container morphism $h =
t \lhd q = \cquo{\tau}$ between $C$ and $C'$ is also a directed
container morphism between $E$ and $E'$.

The counit and comultiplication $\eps$ and $\de$ of the comonad
$\dcsem{E}$ induce container morphisms $h^\eps : C \to \cId$ and
$h^\de : C \to C \cComp C$ by $h^\eps = t^\eps \lhd q^\eps = \cquo{\ee
  \comp \eps}$, $h^\de = t^\de \lhd q^\de = \cquo{\mm \comp
  \de}$. Similarly $\eps'$ and $\de'$ give us container morphisms
$h^{\eps'} : C' \to \cId$ and $h^{\de'} : C' \to C' \cComp C'$ 
by $h^{\eps'} = t^{\eps'} \lhd q^{\eps'} = \cquo{\ee \comp \eps'}$,
$h^{\de'} = t^{\de'} \lhd q^{\de'} = \cquo{\mm \comp \de'}$.

Let us express $h^\eps$ and $h^\de$ directly in terms of $\dn$, $\o$,
$\pl$.

First, from the definitions of $\eps$, $\ee$ we get
\[
h^\eps = \cquo{\ee \comp \eps} = 
 \cquo{\lambda (s, v).\, (\zt, \lambda \zt.\, v\, (\o\, \ia{s}))}
\]
The definition of $\cquo{-}$ further gives us
\begin{eqnarray*}
t^\eps\, s & = & \zt \\
q^\eps\, \ia{s}\, \zt & = & \o\, \ia{s}
\end{eqnarray*}

Second, the definitions of $\de$, $\mm$ dictate that
\[
h^\de = \cquo{\mm\, \ia{C,C} \comp \de} = 
  \cquo{\lambda (s, v).\, (s, \lambda p.\, s \dn p), 
         \lambda (p, p').\, v\, (p \pl \ia{s}\, p')}
\]
The definition of $\cquo{-}$ allows us to infer that
\begin{eqnarray*}
t^\de\, s & = & (s, \lambda p.\, s \dn p) \\
q^\de\, \ia{s}\, (p, p') & = & p \pl \ia{s}\, p' 
\end{eqnarray*}
Analogous direct expressions in terms of $\dnp$, $\op$, $\plp$ hold
for $h^{\eps'}$, $h^{\de'}$.

Now, using $h^\eps, h^\de, h^{\eps'}, h^{\de'}$ above, we can
repackage the two comonad morphism laws for $\tau = \csem{h}$ in terms of
container interpretations as depicted in the following two diagrams.
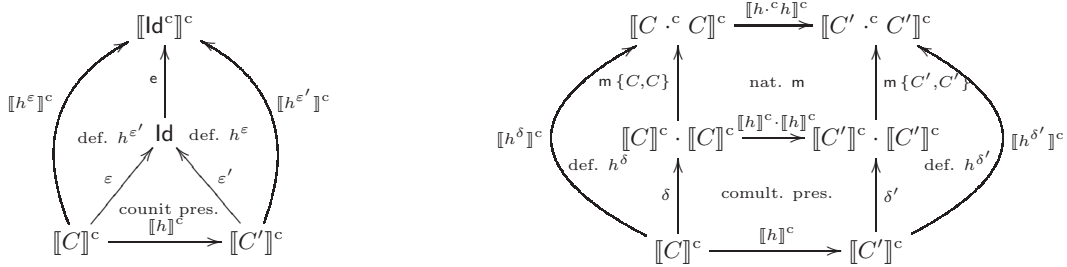
\[
\footnotesize
\xymatrix@C=0.8em@R=3em{
& \csem{\cId} \\
& \Id\ar[u]^{\ee} \ar@{}[l]|(0.6){\textrm{def. $h^{\eps'}$}} \ar@{}[r]|(0.6){\textrm{def. $h^\eps$}} & \\
\csem{C} \ar[rr]^{\csem{h}} \ar[ur]^{\eps} \ar@/^2pc/[uur]^{\csem{\heps}}  \ar@{}[ur]_(0.4){\textrm{counit pres.}} 
& & \csem{C'} \ar[ul]_{\eps'} \ar@/_2pc/[uul]_{\csem{\hepsp}} 
}
\hspace*{2cm}
\xymatrix@C=2.5em@R=3em{
& \csem{C \cComp C}\ar[r]^{\csem{h \cComp h}} 
  & \csem{C' \cComp C'}  \\
& \csem{C} \Comp \csem{C} \ar[r]^{\csem{h} \Comp \csem{h}}\ar[u]^{\mm\, \ia{C, C}} \ar@{}[ur]|{\textrm{nat. $\mm$}} \ar@{}[dr]|{\textrm{comult. pres.}} \ar@{}[dl]_(0.35){\textrm{def. $\hde$}}	
 & {\csem{C'} \Comp \csem{C'}}\ar[u]_{\mm\, \ia{C',C'}} \ar@{}[dr]^(0.35){\textrm{def. $\hdep$}}\\
& \csem{C}\ar[r]^{\csem{h}}\ar[u]^\de \ar@/^4pc/[uu]^{\csem{\hde}}
 & \csem{C'}\ar[u]_{\de'} \ar@/_4pc/[uu]_{\csem{\hdep}} 
     &
}
\]

Going a step further, we can quote these two diagrams to get their
reformulations in terms of containers, resulting in the two diagrams
below.
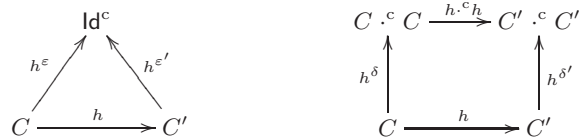
\[
\footnotesize
\xymatrix@C=1.5em@R=3em{
& \cId \\
C \ar[rr]^{h} \ar[ur]^{\heps} & & C' \ar[ul]_{\hepsp} 
}
\hspace*{2cm}
\xymatrix@C=2.5em@R=3em{
{C \cComp C} \ar[r]^{h \cComp h} & {C' \cComp C'}\\
C \ar[r]^{h} \ar[u]^{\hde}  & C' \ar[u]_{\hdep}
}
\]

We are now in a position to prove that $h = \cquo{\tau}$ satisfies
directed container morphism laws.

From the counit preservation law by going clockwise we get that
\[
\footnotesize
\xymatrix@C=4em@R=1em{
s \ar@{|->}[r]& \teps s\\
C \ar[r]^{h^{\eps}} & \cId\\
\qeps\, \ia{s}\, \zt & \ar@{|->}[l] \zt
}
\]
and by going counter-clockwise
\[
\footnotesize
\xymatrix@C=4em@R=1em{
s \ar@{|->}[r] & t s \ar@{|->}[r] & \tepsp (t\, s)\\
C \ar[r]^{h} & C' \ar[r]^{h^{\eps'}} & \cId\\
q\, \ia{s}\, (\qepsp\, \ia{t\, s}\, \zt) & \qepsp\, \ia{t\, s}\, \zt \ar@{|->}[l] & \ar@{|->}[l] \zt
}
\]
which gives us the second directed container morphism law:  
\[
\o\, \ia{s} 
= \qeps\, \ia{s}\,  \zt
= q\, \ia{s}\, (\qepsp\, \ia{t\, s}\, \zt)
= q\, \ia{s}\, (\o'\, \ia{t\, s})
\]

Clockwise traversal of the comultiplication preservation law gives us that
\[
\footnotesize
\xymatrix@C=2.5em@R=1em{
s \ar@{|->}[r] 
& \tde\, s \ar@{|->}[r] 
  & *\txt{$(t\, (\fst\, (\tde\, s)), $\\$\lambda p.\, t\, (\snd\, (\tde\, s) $\\$(q\, \ia{\fst\, (\tde\, s)}\, p)))$}\\
C \ar[r]^{h^\de}
& C \cComp C \ar[r]^{h \cComp h} 
  & C' \cComp C'\\
*\txt{$\qde \ia{s} $\\$(q\,\ia{\fst\, (\tde\, s)}\, p, $\\$q\, \ia{\snd\, (\tde\, s)\, (q\, \ia{\fst\, (\tde\, s)}\, p)}\, p')$} 
& *\txt{$(q\, \ia{\fst\, (\tde\, s)}\, p, $\\$q \ia{\snd\, (\tde\, s)\, (q\, \ia{\fst\, (\tde\, s)}\, p)}\, p')$} \ar@{|->}[l] 
  & (p,p') \ar@{|->}[l]
}
\]
and counter-clockwise traversal that
\[
\footnotesize
\xymatrix@C=2.5em@R=1em{
s \ar@{|->}[r] 
& t\, s \ar@{|->}[r] 
  & *\txt{$\tdep\, (t\, s)$}\\ 
C \ar[r]^{h} 
& C' \ar[r]^{h^{\de'}} 
  & C' \cComp C'\\
q\, \ia{s}\, (\qdep\, \ia{t\, s}\, (p,p')) 
& \qdep\, \ia{t\, s}\, (p,p') \ar@{|->}[l] 
  & (p,p') \ar@{|->}[l]
}
\]
from where we can derive both the first and the third
directed container morphism laws:
\[
t\, (s \dn q\, \ia{s}\, p)
= t\, (\snd\, (\tde\, s)\, (q\, \ia{s}\, p))
= t\, (\snd\, (\tde\, s)\, (q\, \ia{\fst\, (\tde\, s)}\, p))
\]\[= \snd\, (\tdep\, (t\, s))\, p
= t\, s \dnp p
\]
\[
q\, \ia{s}\, p \pl \ia{s}\, q\, \ia{s \dn q\, \ia{s}\, p}\, p'
=\qde \ia{s}\, (q\, \ia{s}\, p, q\, \ia{\snd\, (\tde\, s)\, (q\, \ia{s}\, p)}\, p')
\]\[=\qde \ia{s}\, (q\, \ia{\fst\, (\tde\, s)}\, p, q\, \ia{\snd\, (\tde\, s)\, (q\, \ia{\fst\, (\tde\, s)}\, p)}\, p')
\]\[=q\, \ia{s}\, (\qdep\, \ia{t\, s}\, (p,p'))
=q\, \ia{s}\, (p \pl' \ia{t\, s}\, p')\rlap{\hbox to 110 pt{\hfill\qEd}}
\]

\subsection*{Proof of Proposition \ref{prop:comonad2dcontainer}}

\proof
We need to verify that $(C, \dn, \o, \pl)$ satisfies the
directed container laws and can assume that $(D, \eps, \de)$ satisfies
the comonad laws.

The comonad laws can be rewritten in terms of container
interpretations as outlined in the following commuting diagrams:
\[
\footnotesize
\xymatrix@C=3em@R=3em{
\csem{C \cComp C} \ar[r]^{\csem{C \cComp \heps}} 
& \csem{C \cComp \cId} \ar@/^2.5pc/[ddr]^{\csem{\rho}}\\
\ar@{}[ur]|{\textrm{nat.\, $\mm$}} 
  & \csem{C} \Comp \csem{\cId}\ar[u]_{\mm\, \ia{C,\Id}}  \ar@{}[r]_(0.65){\textrm{mon.\ f.\ r.\ unit}}
    & \\
\csem{C} \Comp \csem{C} \ar[r]^{\csem{C} \Comp \eps} \ar[uu]^{\mm\, \ia{C,C}} \ar@/^1pc/[ur]^{\csem{C} \Comp \csem{\heps}}  \ar@{}[ur]|{\textrm{def. $\heps$}} \ar@{}[uu]^(0.2){\textrm{def. $\hde$}} 
& \csem{C} \Comp \Id \ar@{=}[r] \ar[u]_{\csem{C} \Comp \ee}  
  & \csem{C}\\
\csem{C} \ar[u]^{\de} \ar@/^3pc/[uuu]^{\csem{\hde}} \ar@{=}[urr] \ar@{}[ur]^{\textrm{com. r.\ counit}} 
}
\xymatrix@C=3em@R=3em{
& \csem{\cId \cComp C} \ar@/_2.5pc/[ddl]_{\csem{\lambda}} & \csem{C \cComp C} \ar[l]_{\csem{\heps \cComp C}} \\
& \ar@{}[l]^(0.65){\textrm{mon.\ f.\ l.\ unit}} \ar@{}[ur]|{\textrm{nat. $\mm$}} \csem{\cId} \Comp \csem{C} \ar[u]^{\mm\, \ia{\Id, C}}\\
\csem{C} 
& \Id \Comp \csem{C} \ar@2{-}[l] \ar[u]^{\ee \Comp \csem{C}}  
  & \csem{C} \Comp \csem{C} \ar[l]_{\eps \Comp \csem{C}} \ar[uu]_{\mm\, \ia{C,C}} \ar@/_1pc/[ul]_{\csem{\heps} \Comp \csem{C}} \ar@{}[ul]|{\textrm{def. $\heps$}} \ar@{}[uu]_(0.2){\textrm{def. $\hde$}} \\
& & \csem{C} \ar[u]_{\de} \ar@2{-}[ull] \ar@{}[ul]_{\textrm{com. l. counit}} \ar@/_3pc/[uuu]_{\csem{\hde}}
}
\]

\[
\footnotesize
\xymatrix@C=2.5em@R=4em{
\csem{C} \ar@{}[drr]_{\textrm{com. coass.}} \ar[rr]_{\de} \ar[d]^{\de} \ar@/^2pc/[rrr]^{\csem{\hde}}_{\textrm{def. $\hde$}} \ar@/_3pc/[dd]_{\csem{\hde}} 
  & & \csem{C} \Comp \csem{C}  \ar[r]^{\mm\, \ia{C,C}} \ar[d]_{\de \Comp \csem{C}} \ar@/^0.5pc/[dr]^{\csem{\hde} \Comp \csem{C}}  \ar@{}[dr]_{\textrm{def. $\hde$}}
      & \csem{C \cComp C} \ar@/^3pc/[dd]^{\csem{\hde \cComp C}} \\
\csem{C} \Comp \csem{C} \ar@{}[u]^{\textrm{def. $\hde$}} \ar@{}[dr]^{\textrm{def. $\hde$}} \ar[d]^{\mm\, \ia{C,C}} \ar[r]^-{\csem{C} \Comp \de} \ar@/_0.5pc/[dr]_{\csem{C} \Comp \csem{\hde}} 
  & \csem{C} \Comp (\csem{C} \Comp \csem{C}) \ar[d]^{\csem{C} \Comp \mm\, \ia{C,C}} 
        \ar@{}[drr]|{\textrm{mon. f. ass.}}
    & (\csem{C} \Comp \csem{C}) \Comp \csem{C} \ar@{=}[l] \ar[r]_{\mm\, \ia{C,C} \Comp \csem{C}}     
      & \csem{C \cComp C} \Comp \csem{C} \ar[d]_{\mm\, \ia{C \cComp C, C}} \ar@{}[u]|{\textrm{nat. $\mm$}}\\
\csem{C \cComp C} \ar@/_2pc/[rr]_{\csem{C \cComp \hde}} 
  &  \csem{C} \Comp \csem{C \cComp C} \ar[r]^{\mm\, \ia{C, C \cComp C}} \ar@{}[d]|(0.3){\textrm{nat. $\mm$}}
    & \csem{C \cComp (C \cComp C)}
      &  \csem{(C \cComp C) \cComp C} \ar[l]_{\csem{\alpha}} \\
 & & 
}
\]

Next we quote these three diagrams to get the comonad laws in terms
of containers in the next three commuting diagrams.
\[
\footnotesize
\xymatrix@C=3.5em@R=3em{
C \cComp C \ar[r]^-{C \cComp \heps} & C \cComp \cId \ar[r]^{\rho}  & C\\
C \ar[u]^{\hde} \ar@{=}[urr]
}
\quad
\xymatrix@C=3.5em@R=3em{
C & \cId \cComp C \ar[l]_{\lambda} & C \cComp C \ar[l]_-{\heps \cComp C} \\
& & C \ar[u]_{\hde} \ar@{=}[ull]
}
\]
\[
\footnotesize
\xymatrix@C=2em@R=3em{
C \ar[rr]^{\hde} \ar[d]_{\hde} 
  & 
    & C \cComp C \ar[d]^{\hde \cComp C} \\
C \cComp C \ar[r]^-{C \cComp \hde} 
  &  C \cComp (C \cComp C)
    & (C \cComp C) \cComp C \ar[l]_-{\alpha}
}
\]

From the comonad right counital law we get by going clockwise
\[
\footnotesize
\xymatrix@C=2.5em@R=1em{
s \ar@{|->}[r] & \tde s \ar@{|->}[r] 
& *\txt{$(\fst\, (\tde\, s),$\\$\lambda\ \_.\, \zt)$} \ar@{|->}[r] 
  & \fst\, (\tde\, s)\\
C \ar[r]^{\hde} 
& C \cComp C \ar[r]^{C \cComp \heps} 
  & C \cComp \cId \ar[r]^{\rho} & C\\
*\txt{$\qde \ia{s}\, (p, $\\$\qeps\, \ia{\snd\, (\tde\, s)\, p}\, \zt)$} 
& *\txt{$(p, $\\$\qeps \{\snd\, (\tde\, s)\, p\}\, \zt)$} \ar@{|->}[l] 
  & (p,\zt) \ar@{|->}[l] & p \ar@{|->}[l]
}
\]
from where it follows that $\de$ preserves the shape
of the given data-structure as the outer shape of the composite
data-structure returned and that the third directed container law holds:
\[
s 
= \fst\, (\tde\, s)
\]
\[
p 
= \qde \ia{s} (p, \qeps \ia{\snd\, (\tde s)\, p}\, \zt) 
= p \pl \ia{s}\, \o\, \ia{s \dn p}
\]

Similarly, from the comonad left counital law we get by going counter-clockwise
\[
\footnotesize
\xymatrix@C=1.3em@R=1em{
s \ar@{|->}[r] 
& \tde s \ar@{|->}[r] 
  & *\txt{$(\zt, \lambda \zt.\, \snd\, (\tde\, s) $\\$(\qeps\, \ia{\fst\, (\tde\, s)}))$} \ar@{|->}[r] 
    & *\txt{$\snd\, (\tde\, s) $\\$(\qeps\, \ia{\fst\, (\tde\, s)}\, \zt)$}\\
C \ar[r]^{\hde} 
& C \cComp C \ar[r]^{\heps \cComp C} 
  & \cId \cComp C \ar[r]^{\lambda} 
    & C\\
*\txt{$\qde\, \ia{s} $\\$(\qeps\, \ia{\fst\, (\tde\, s)}\, \zt, p)$} 
& *\txt{$(\qeps\, \ia{\fst\, (\tde\, s)}\, \zt, $\\$p)$} \ar@{|->}[l] 
  & (\zt,p) \ar@{|->}[l] 
    & p \ar@{|->}[l]
}
\]
from where the first and fourth directed container laws follow:
\[
s 
= \snd\, (\tde\, s)\, (\qeps\, \ia{\fst\, (\tde\, s)}\, \zt) 
= \snd\, (\tde\, s)\, (\qeps\, \ia{s}) 
= s \dn \o\, \ia{s}
\]
\[
p = 
\qde \ia{s}\, (\qeps\, \ia{\fst\, (\tde\, s)}\, \zt, p) 
= \qde \ia{s}\, (\qeps\, \ia{s}\, \zt, p)
= \o\, \ia{s} \pl \ia{s}\, p
\]

The last two directed container laws are derivable from the comonad
coassociativity law. By going clockwise we get 
\[
\footnotesize
\xymatrix@C=2.8em@R=1em{
s \ar@{|->}[r] 
& \tde s \ar@{|->}[r] 
  & *\txt{$(\tde\, (\fst\, (\tde\, s)), $\\$\snd\, (\tde\, s) \comp $\\$(\qde\, \ia{\fst\, (\tde\, s)}))$} \ar@{|->}[r] 
    & *\txt{$(\fst\, (\tde\, (\fst\, (\tde\, s))), $\\$(\lambda p.\, \snd\, (\tde\, (\fst\, (\tde s)))\, p,$\\$\lambda p'.\, \snd\, (\tde\, s) $\\$(\qde\, \ia{\fst\, (\tde\, s)}\, (p , p'))))$}\\
C \ar[r]^{\hde} 
& C \cComp C \ar[r]^{\hde \cComp C} 
  & (C \cComp C) \cComp C \ar[r]^{\alpha} 
    & C \cComp (C \cComp C)\\
*\txt{$(\qde \ia{s} $\\$(\qde\, \ia{\fst\, (\tde\, s)} $\\$(p,p'),p'')$} 
& *\txt{$(\qde \ia{\fst\, (\tde\, s)} $\\$(p,p'), p'')$} \ar@{|->}[l] 
  & ((p,p'),p'') \ar@{|->}[l] 
    & (p , (p' , p'')) \ar@{|->}[l]
}
\]
and by going counter-clockwise we get
\[
\footnotesize
\xymatrix@C=3em@R=1em{
s \ar@{|->}[r] & \tde s \ar@{|->}[r] 
& *\txt{$(\fst\, (\tde\, s), $\\$(\lambda p.\, \fst\, (\tde\, (\snd\, (\tde\, s)\, p)), $\\$\lambda p'.\, \snd\, (\tde\, (\snd\, (\tde s)\, p))\, p'))$}\\
C \ar[r]^{\hde} 
& C \cComp C \ar[r]^{C \cComp \hde} & C \cComp (C \cComp C)\\
*\txt{$(\qde\, \ia{s}\, (p, $\\$\qde\, \ia{\snd\, (\tde\, s)\, p}\, (p',p'')))$}
& *\txt{$(p, $\\$\qde\, \ia{\snd\, (\tde\, s)\, p}\, (p',p''))$} \ar@{|->}[l] & (p,(p',p'')) \ar@{|->}[l]
}
\]
from where the second and fifth directed container laws follow
\[
s \dn (p \pl p') 
= \snd\, (\tde\, s)\, (\qde\, \{s\}\, (p , p')) 
= \snd\, (\tde\, s)\, (\qde\, \{\fst\, (\tde\, s)\}\, (p , p')) 
= \]\[ \snd\, (\tde\, (\snd\, (\tde\, s)\, p))\, p' 
= (s \dn p) \dn p'
\]
\[
(p \pl \ia{s}\, p') \pl \ia{s}\, p'' 
= \qde\, \ia{s}\, (\qde\, \ia{s}\, (p',p''),p'') 
= \qde\, \ia{s}\, (\qde\, \ia{\fst\, (\tde\, s)}\, (p',p''),p'') 
= \]\[\qde\, \ia{s}\, (p, \qde\, \ia{\snd\, (\tde\, s)\, p}\, (p',p'')) 
= p \pl \ia{s}\, (p' \pl \ia{s \dn p}\, p'')\rlap{\hbox to 75 pt{\hfill\qEd}}
\]

\subsection*{Proof of Proposition \ref{prop:pullbacklaw2}}

\proof
By interpreting the given directed container $(C,\dn,\o,\pl)$ we get a
comonad $(D,\eps,\de)$ whereby $D=\csem{C}$, $\eps\, (s,v) = v\, (\o\,
\ia{s})$ and $\de\, (s,v) = (s , \lambda p.\, (s \dn p,
\lambda p'.\, v\, (p \pl \ia{s}\, p')))$.

From the comonad, we get a directed container $(C,\dn',\o',\pl') =
\lceil (D,\eps,\de), C \rceil$ by taking $s \dn' p = \snd\, (t^\de\,
s)\, p$, $\o'\, \ia{s} = q^\eps\, \ia{s}\, \zt$, $p \pl' \ia{s}\, p' =
q^\de\, \ia{s}\, (p, p')$. 

This directed container must be equal to the original directed
container $(C,\dn,\o,\pl)$, i.e., we need to prove that $s \dn' p = s
\dn p$ and $\o'\, \ia{s} = \o\, \ia{s}$ and $p \pl' p' = p \pl p'$.

By the definitions of $\ee$, $\mm$, $\cquo{-}$, for the container
morphisms $\teps \lhd \qeps = \cquo{\ee \comp \eps}$ and $\tde \lhd
\qde = \cquo{\mm\, \comp\, \de} $ we have that
\[\teps\, s = \zt\]
\[\qeps\, \ia{s}\, \zt = \eps\, (s, \id)\]
\[\tde\, s = (\fst\, (\de\,(s, \id)) , \lambda  p.\, \fst\, (\snd\, (\de\, (s, \id))\, p))\]
\[\qde\, \ia{s}\, (p,p') = \snd\, (\snd\, (\de\, (s,\id))\, p)\, p'\]

Using the definitions of $\dn'$, $\o'$, $\pl'$, $\eps$, $\de$, we calculate:
\[
s \dn' p = \snd\, (\tde\, s)\, p
= \fst\, (\snd\, (\de\, (s, \id))\, p))
= s \dn p
\]
\[
\o'\, \ia{s} = \qeps\, \ia{s}\, \zt
= \eps\, (s, \id)
= \o\, \ia{s} 
\]
\[
p \pl' \ia{s}\, p' = \qde\, \ia{s}\, (p,p')
  = \snd\, (\snd\, (\de\, (s,\id))\, p)\, p'
  = p \pl \ia{s}\, p'\rlap{\hbox to 67 pt{\hfill\qEd}}
\]

\subsection*{Proof of Proposition \ref{prop:pullbacklaw1}}

\proof 
The comonad $(D, \eps, \de)$ induces a directed container $(S \lhd
P,\dn,\o,\pl) = \lceil (D, \eps, \de), S \lhd P\rceil$ whereby
\[s \dn p = \fst\, (\snd\, (\de\, (s, \id))\, p)\]
\[\o\, \ia{s} = \eps\, \{P\, s\}\, (s, \id)\]
\[p \pl \ia{s}\, p' = \snd\, (\snd\, (\de\, (s,\id))\, p)\, p'\]

By interpreting this directed container, we get a comonad
$(D', \eps', \de') = \dcsem{S \lhd P,{\dn},\o,{\pl}}$ by taking 
$\eps'\, (s,v) = v\, (\o\,
\ia{s})$ and $\de'\, (s,v) = (s , \lambda p.\, (s \dn p,
\lambda p'.\, v\, (p \pl \ia{s}\, p')))$.\newpage

This comonad must equal $(D,\eps,\de)$, i.e., we need to prove 
that $D' = D$ and $\eps'\, \ia{X}\, (s,v) = \eps\, \ia{X}\, (s,v)$ and $\de'\,
\ia{X}\, (s,v)$ = $\de\, \ia{X}\, (s,v)$.

First of all, from the definition of directed container
interpretation, we know that the underlying functors are equal: $D' =
\csem{S \lhd P} = D$.

Using the definitions of $\eps'$, $\de'$, $\dn$, $\o$, $\pl$ we can calculate
\[
\eps'\, \ia{X} (s,v) 
= v\, (\o\, \ia{s}) 
= v\, (\qeps\, \ia{s}\, \zt) 
= v\, (\eps\, \ia{P\, s}\, (s, \id))
\] 
\[
\de'\, \ia{X} (s,v) 
= (s, \lambda p.\, s \dn p\, ,\, \lambda p'.\, v\, (p \pl\ia{s}\, p')) 
= \] 
\[
= (s, \lambda p.\, (\fst\, (\snd\, \de\, \{P\, s\}\, (s, \id)\, p), \lambda p'.\, v\, (\snd\, (\snd\, \de \, \ia{P\, s}\,  (s,\id)\, p)\, p')))
\]

Now, because of naturality of $\eps$ and $\de$ expressed in the diagrams
\[
\footnotesize
\xymatrix@C=5em@R=1.5em{
(s,\id) \ar@{|->}[r] \ar@{|->}@/_5pc/[dddd] & \eps\, \ia{P\, s}\, (s,\id) \ar@{|->}@/^5pc/[dddd] \\
\Sigma s : S . P\, s \to P\, s \ar[r]^-{\eps\, \ia{P\, s}} \ar[dd]_{\lambda (s,v').\, (s,v \comp v')} & P\, s \ar[dd]^{v}\\
& \\
\Sigma s : S . P\, s \to X \ar[r]_-{\eps\, \ia{X}} & X\\
(s,v) \ar@{|->}[r] & \eps\, \ia{X}\, (s,v) = v\, (\eps\, \ia{P\, s}\, (s,\id))
}
\]

\[
\footnotesize
\xymatrix@C=5em@R=1.5em{
(s,\id) \ar@{|->}[r] \ar@{|->}@/_6pc/[dddd] & \de\, \{P s\}\, (s,\id) \ar@{|->}@/^12.5pc/[dddd] \\
\Sigma s : S . P\, s \to P\, s \ar[r]^-{\de\, \{P\, s\}} \ar[dd]_{\lambda (s,v').\, (s,v \comp v')} & \Sigma s : S . P\, s \to \Sigma s' : S . P\, s' \to P\, s \ar[dd]^{\lambda (s,v').\, (s, \lambda p.\, (\fst\, (v'\, p), v \comp \snd\, (v'\, p)))}\\
& \\
\Sigma s : S . P\, s \to X \ar[r]_-{\de\, \ia{X}} & \Sigma s : S . P\, s \to \Sigma s' : S . P\, s' \to X \\
(s,v) \ar@{|->}[r] & *\txt{$\de\, \ia{X}\, (s,v) = $\\$(\fst\, (\de\, \ia{P\, s}\, (s,\id)), $\\$\lambda p.\, (\fst\, (\snd\, \de\, \ia{P\, s}\, (s, \id)\, p), $\\$\lambda p'.\, v\, (\snd\, (\snd\, \de \, \ia{P\, s}\,  (s,\id)\, p)\, p')))$}
}
\]
it is evident that the counit and comultiplication of $(D,\eps,\de)$ and $(D',\eps',\de')$ are equal:
\[\eps'\, \ia{X}\, (s,v) = \eps\, \ia{X}\, (s,v)\] 
\[\de'\, \ia{X}\, (s,v) = \de\, \ia{X}\, (s,v)\eqno{\qEd}\]


\section{Proofs for Section~\ref{sec:constructions}}

\subsection*{Proof of Proposition~\ref{prop:prod}}\hfill

We must show that the definitions yield a strict directed container
that is a product of two given strict directed containers in the
category of all directed containers.

We first check that $E$ is a strict directed container.\newpage

\begin{lem}
The data $\dntpl$ and $\pltpl$ equip the container $S \lhd P^+$ with a
strict directed container structure.
\end{lem}

\proof Auxiliary statements $s \dnpll (p \plpll p') = (s \dnpll p)
\dntpl p'$ and $s \dnplr (p \plplr p') = (s \dnplr p) \dntpl p'$ for
law 1, by mutual induction on the two $p$s. We show only the cases for
the first auxiliary statement; those of the second are symmetric.

\noindent Case $p=(p_0,\nothing)$, $p' = \inl\,(p_0', \nothing)$:
\[
\begin{array}{cl}
&(s_0, v_0) \dnpll ((p_0, \nothing) \plpll
  \inl\,(p_0', \nothing))\\
=&\quad\{\textrm{definitions of $\dnpll$, $\plpll$}\}\\
&((s_0\dnpla (p_0 \plpla p_0'), 
   \lambda p.\,v_0\,((p_0 \plpla p_0')\plpla p)),
 v_0\,(p_0 \plpla p_0'))\\
=&\quad\{\textrm{strict directed container laws 1, 2} \}\\
&(((s_0\dnpla p_0) \dnpla p_0', \lambda p.\,v_0\,(p_0\plpla (p_0'\plpla p))),
  v_0\,(p_0\plpla p_0'))\\
=&\quad\{\textrm{definitions of $\dnpll$, $\dntpl$} \}\\
&((s_0, v_0) \dnpll (p_0, \nothing)) \dntpl
  \inl\,(p_0', \nothing)
\end{array}
\]
Case for $p=(p_0, \nothing)$, $p'=\inl\,(p_0', \just\,p_1')$:
\[
\begin{array}{cl}
&(s_0, v_0) \dnpll ((p_0, \nothing) \plpll \inl\,(p_0', \just\, p_1'))\\
=&\quad\{\textrm{definitions of $\dnpll$, $\plpll$}\}\\
&v_0\, (p_0 \plpla p_0') \dnplr p_1'\\
=&\quad\{\textrm{definitions of $\dnpll$, $\dntpl$} \}\\
&( (s_0,v_0)\dnpll (p_0, \nothing))\dntpl 
  \inl\,(p_0',\just\, p_1')
\end{array}
\]
Case $p = (p_0, \nothing)$, $p' = \inr\,p'$:
\[
\begin{array}{cl}
& (s_0, v_0) \dnpll((p_0,\nothing) \plpll \inr\,p')\\
=&\quad\{\textrm{definitions of $\dnpll$, $\plpll$}\}\\
&v_0\, p_0 \dnplr p'\\
=&\quad\{\textrm{definitions of $\dnpll$, $\dntpl$} \}\\
&((s_0,v_0) \dnpll (p_0,\nothing))\dntpl \inr\,p'
\end{array}
\]
Case $p= (p_0,\just\,p_1)$:
\[
\begin{array}{cl}
&(s_0,v_0) \dnpll ((p_0,\just\, p_1)\plpll p')\\
=&\quad\{\textrm{definitions of $\dnpll$, $\plpll$}\}\\
& v_0\, p_0 \dnplr (p_1 \plplr p') \\
=&\quad\{\textrm{inductive hypothesis for $p_1$}\}\\
& (v_0\, p_0 \dnplr p_1) \dntpl p')\\
=&\quad\{\textrm{definition of $\dnpll$}\}\\
&((s_0,v_0) \dnpll (p_0,\just\,p_1)) \dntpl\,p'
\end{array}
\]

\noindent Strict directed container law 1. Case $s = (s_0, s_1)$, $p = \inl\,
p$:
\[
\begin{array}{cl}
&(s_0,s_1) \dnpl (\inl\, p \plpl p')\\
=&\quad\{\textrm{definition of $\plpl$}\}\\
&(s_0, s_1) \dnpl \inl\, (p \plpll p')\\
=&\quad\{\textrm{definition of $\dnpl$}\}\\
&s_0 \dnpll (p \plpll p')\\
=&\quad\{\textrm{first aux. statement}\}\\
&(s_0 \dnpll p) \dnpl p'\\
=&\quad\{\textrm{definition of $\dnpl$}\}\\
&((s_0, s_1) \dnpl \inl\, p) \dnpl p'
\end{array}
\]
Case $p = \inr\, p$ is symmetric.

\noindent Auxiliary statements $(p\plpll p')\plpll p''=
p\plpll(p'\pltpl p'')$ and $(p\plplr p')\plplr p''= p\plplr(p'\pltpl
p'')$ for law 2, by mutual induction on the two $p$s. We show only the
cases of the first statement.  Case $p=(p_0, \nothing)$,
$p'=\inl\,(p_0',\nothing)$, $p'' = \inl\,(p_0'', p_1'')$:
\[
\begin{array}{cl}
&((p_0,\nothing)\plpll\inl\,(p_0',\nothing))\plpll \inl\,(p_0'',p_1'')\\
=&\quad\{\textrm{definition of $\plpll$}\}\\
&((p_0\plpla p_0')\plpla p_0'',p_1'')\\
=&\quad\{\textrm{strict direct container law 2}\}\\
&(p_0\plpla(p_0'\plpla p_0''),p_1'')\\
=&\quad\{\textrm{definitions of $\plpll$, $\pltpl$}\}\\
&(p_0,\nothing)\plpll(\inl\,(p_0',\nothing)\pltpl \inl\,(p_0'',p_1''))
\end{array}
\]
Case $p=(p_0, \nothing)$, $p'=\inl\,(p_0',\nothing)$, $p'' =
\inr\,p''$:
\[
\begin{array}{cl}
&((p_0,\nothing)\plpll\inl\,(p_0',\nothing))\plpll \inr\,p''\\
=&\quad\{\textrm{definition of $\plpll$}\}\\
&(p_0\plpla p_0',\just\,p'')\\
=&\quad\{\textrm{definitions of $\plpll$, $\pltpl$}\}\\
&(p_0,\nothing)\plpll(\inl\,(p_0',\nothing)\pltpl \inr\,p'')
\end{array}
\]
Case $p=(p_0, \nothing)$, $p'=\inl\,(p_0',\just\,p_1')$:
\[
\begin{array}{cl}
&((p_0,\nothing)\plpll \inl\,(p_0',\just\,p_1'))\plpll p''\\
=&\quad\{\textrm{definition of $\plpll$}\}\\
&(p_0 \plpla p_0',\just\,(p_1'\plplr p''))\\
=&\quad\{\textrm{definitions of $\plpll$, $\pltpl$}\}\\
&(p_0,\nothing)\plpll(\inl\,(p_0',\just\,p_1')\pltpl p'')
\end{array}
\]
Case $p = (p_0, \nothing)$, $p' = \inr\, p'$:
\[
\begin{array}{cl}
&((p_0,\nothing)\plpll \inr\, p')\plpll p''\\
=&\quad\{\textrm{definition of $\plpll$}\}\\
&(p_0, \just\, (p' \plplr p'')) \\
=&\quad\{\textrm{definitions of $\plpll$, $\pltpl$}\}\\
&(p_0,\nothing)\plpll(\inr\, p'\pltpl p'')\\
\end{array}
\]
Case $p= (p_0, \just\,p_1)$:
\[
\begin{array}{cl}
&((p_0,\just\,p_1)\plpll p')\plpll p''\\
=&\quad\{\textrm{definition of $\plpll$}\}\\
&(p_0,\just\,((p_1\plplr p')\plplr p''))\\
=&\quad\{\textrm{inductive hypothesis for $p_1$}\}\\
&(p_0,\just\,(p_1\plplr(p'\pltpl p'')))\\
=&\quad\{\textrm{definition of $\plpll$}\}\\
&(p_0,\just\,p_1)\plpll(p'\pltpl p'')\\
\end{array}
\]

\noindent Strict directed container law 2. Case $p = \inl\, p$:
\[
\begin{array}{cl}
& (\inl\, p \plpl p') \plpl p''\\
=&\quad\{\textrm{definition of $\plpl$}\}\\
& \inl\, (p \plpll p') \plpl p''\\
=&\quad\{\textrm{definition of $\plpl$}\}\\
& \inl\, ((p \plpll p') \plpll p')\\
=&\quad\{\textrm{first aux. statement}\}\\
& \inl\, (p \plpll (p' \plpl p'')\\
=&\quad\{\textrm{definition of $\plpl$}\}\\
& \inl\, p \plpl (p' \plpl p'')
\end{array}
\]
Case $p = \inr\, p$ is symmetric.
\qed

To check that $E$ is a product of $E_0$ and $E_1$ we can either
verify it directly that it satisfies the required universal property
or prove that it interprets to a product of the interpreting
comonads. Here we have chosen to pursue the first route. 

For $E$ to be a product of $E_0$ and $E_1$, it must come with
directed container morphisms $\pi_0 = t^{\pi_0} \lhd
q^{\pi_0} : E_0 \to E$, $\pi_1 = t^{\pi_1} \lhd q^{\pi_1} : E_1 \to
E$. We claim that they can be defined by

\begin{itemize}
\item $t^{\pi_0} : S \to S_0$\\
          $t^{\pi_0}\, ((s_0 , v_0) , (s_1 , v_1)) = s_0$
\item $t^{\pi_1} : S \to S_1$\\
          $t^{\pi_1}\, ((s_0 , v_0) , (s_1 , v_1)) = s_1$
\item $q^{\pi_0} : \Pi \ia{s : S} .\, P_0\, (t^{\pi_0}\, s) \to P\, s$ \\
          $q^{\pi_0}\, \nothing = \nothing$\\
          $q^{\pi_0}\, (\just\, p) = \just\, (\inl\, (p , \nothing))$
\item $q^{\pi_1} : \Pi \ia{s : S} .\, P_1\, (t^{\pi_1}\, s) \to P\, s$ \\
          $q^{\pi_1}\, \nothing = \nothing$\\
          $q^{\pi_1}\, (\just\, p) = \just\, (\inr\, (p , \nothing))$
\end{itemize}

\noindent Moreover, any directed container $E' = (S' \lhd P', {\dnp}, \op,
{\plp})$ with two directed container morphisms $f_0 = t^{f_0} \lhd
q^{f_0} : E' \to E_0$ and $f_1 = t^{f_1} \lhd q^{f_1}: E' \to E_1$
must jointly determine a unique directed container morphism $f = t^f
\lhd q^f : E' \to E$ such that the following two triangles commute.
\[
\tag{$\dag$}
\label{diag:products}
\begin{gathered}
\xymatrix@C=4.3em@R=4em{
& (S' \lhd P' , \dnp , \op , \plp) \ar[dl]_{t^{f_0} \lhd q^{f_0}} \ar[dr]^{t^{f_1} \lhd q^{f_1}} \ar[d]_{t^f \lhd q^f}& \\
(S_0 \lhd P_0 , \dna , \oa , \pla) & \ar[l]^{t^{\pi_0} \lhd q^{\pi_0}} (S \lhd P , \dnt , \ot , \plt) \ar[r]_{t^{\pi_1} \lhd q^{\pi_1}} & (S_1 \lhd P_1 , \dnb , \ob , \plb)
}
\end{gathered}
\]
\noindent We claim that $f$ is given by
\begin{itemize}
\item $t^f : S' \to S$\\
$t^f\, s = (\tfbarzero\,s,\tfbarone\,s)$\\
where\\
$\tfbarzero : S' \to \overline{S_0}$\\
$\tfbarone : S' \to \overline{S_1}$\\
(by mutual corecursion)\\
$\tfbarzero\,s = (\tfzero\, s, \lambda p .\, \tfbarone\, (s \dnp \qfzero\, (\just\, p)))$\\
$\tfbarone\,s = (\tfone\, s, \lambda p .\, \tfbarzero\, (s \dnp \qfone\, (\just\, p)))$\\

\item $q^f : \Pi \ia{s : S'}.\, P\, (t^f\, s) \to P'\, s$\\
$q^f\, \nothing = \op$\\
$q^f\, (\just\,(\inl\,p)) = \qfbarzero\,p$\\
$q^f\, (\just\,(\inr\,p)) = \qfbarone\,p$\\
where \\
$\qfbarzero : \Pi \ia{s : S'}.\, \Pplusbarzero\, (\tfbarzero\, s) \to P'\, s$\\
$\qfbarone : \Pi \ia{s : S'}.\, \Pplusbarone\, (\tfbarone\, s) \to P'\, s$\\
(by mutual recursion) \\
$\qfbarzero\, (p_0 , \nothing) = \qfzero\,(\just\,p_0)$\\
$\qfbarzero\, (p_0 , \just\, p_1) = \qfzero\,(\just\, p_0) \plp \qfbarone p_1$\\
$\qfbarone\, (p_1 , \nothing) = \qfone\,(\just\,p_1)$\\
$\qfbarone\, (p_1 , \just\, p_1) = \qfone\,(\just\, p_1) \plp \qfbarzero p_0$
\end{itemize}

\begin{lem}
The container morphisms $\pi_0$, $\pi_1$ are directed container
morphisms.\footnote{They are in fact strict directed container
  morphisms, but we will not prove this here, as we have not
  defined this concept.}
\end{lem}

\proof We give the proof only for $\pi_0$. The proof for $\pi_1$ is
symmetric.

\noindent Directed container morphism law 1. Case $p = \nothing$:
\[
\begin{array}{cl}
&t^{\pi_0}\,(s \dnt (q^{\pi_0}\, \nothing))\\
=&\quad\{\textrm{definition of $q^{\pi_0}$} \}\\
&t^{\pi_0}\,(s \dnt \nothing)\\
=&\quad\{\textrm{definition of $\dnt$}\}\\
&t^{\pi_0}\,s\\
=&\quad\{\textrm{definition of $\dna$}\}\\
&t^{\pi_0}\,s \dna \nothing
\end{array}
\]
Case $p = \just\, p$:
\[
\begin{array}{cl}
&t^{\pi_0}\,(((s_0 , v_0) , (s_1 , v_1))  \dnt q^{\pi_0}\, (\just\, p))\\
=&\quad\{\textrm{definition of $q^{\pi_0}$} \}\\
&t^{\pi_0}\,(((s_0 , v_0) , (s_1 , v_1)) \dnt \just\, (\inl\, (p , \nothing)))\\
=&\quad\{\textrm{definition of $\dnt$} \}\\
&t^{\pi_0}\,(((s_0 , v_0) , (s_1 , v_1)) \dntpl \inl\, (p , \nothing))\\
=&\quad\{\textrm{definition of $\dntpl$} \}\\
&t^{\pi_0}\,((s_0 , v_0) \dnpll (p , \nothing))\\
=&\quad\{\textrm{definition of $\dnpll$} \}\\
&t^{\pi_0}\,((s_0 \dnpla p , \lambda p'.\, v_0 \,(p \plpla p')) , v_0\, p)\\
=&\quad\{\textrm{definition of $t^{\pi_0}$} \}\\
&s_0 \dnpla p\\
=&\quad\{\textrm{definition of $t^{\pi_0}$} \}\\
&t^{\pi_0}\, ((s_0 , v_0) , (s_1 , v_1)) \dnpla p\\
=&\quad\{\textrm{definition of $\dna$} \}\\
&t^{\pi_0}\, ((s_0 , v_0) , (s_1 , v_1)) \dna \just\, p
\end{array}
\]

\noindent Directed container morphism law 2:
\[
\begin{array}{cl}
&q^{\pi_0}\,\oa\\
=&\quad\{\textrm{definition of $\oa$} \}\\
&q^{\pi_0}\,\nothing\\
=&\quad\{\textrm{definition of $q^{\pi_0}$} \}\\
&\nothing\\
=&\quad\{\textrm{definition of $\ot$} \} \\
&\ot
\end{array}
\]
\noindent Directed container morphism law 3. Case $p = \nothing$:
\[
\begin{array}{cl}
&q^{\pi_0}\,(\nothing \pla p')\\
=&\quad\{\textrm{definition of $\pla$} \}\\
&q^{\pi_0}\, p'\\
=&\quad\{\textrm{definition of $\plt$} \}\\
&\nothing \plt q^{\pi_0}\, p'
\end{array}
\]
\noindent Case $p = \just\, p$, $p' = \nothing$:
\[
\begin{array}{cl}
&q^{\pi_0}\,(\just\, p \pla \nothing)\\
=&\quad\{\textrm{definition of $\pla$} \}\\
&q^{\pi_0}\, (\just\, p)\\
=&\quad\{\textrm{definition of $q^{\pi_0}$} \}\\
&\just\, (\inl\, (p, \nothing))\\
=&\quad\{\textrm{definition of $\plt$} \}\\
&\just\, (\inl\, (p, \nothing)) \plt \nothing \\
=&\quad\{\textrm{definition of $q^{\pi_0}$} \}\\
&q^{\pi_0}\, (\just\, p) \plt \nothing
\end{array}
\]
Case $p = \just\, p$, $p' = \just\, p'$:
\[
\begin{array}{cl}
&q^{\pi_0}\,(\just\, p \pla \just\, p')\\
=&\quad\{\textrm{definition of $\pla$}\}\\
&q^{\pi_0}\,(\just\, (p \plpla p'))\\
=&\quad\{\textrm{definition of $q^{\pi_0}$} \}\\
&\just\, (\inl\, (p \plpla p' , \nothing))\\
=&\quad\{\textrm{definition of $\plpll$}\}\\
&\just\, (\inl\, ((p , \nothing) \plpll \inl\, (p' , \nothing)))\\
=&\quad\{\textrm{definition of $\pltpl$}\}\\
&\just\, (\inl\, (p , \nothing) \pltpl \inl\, (p' , \nothing))\\
=&\quad\{\textrm{definition of $\plt$}\}\\
&\just\, (\inl\, (p , \nothing) \plt \just\, (\inl\, (p' , \nothing))\\
=&\quad\{\textrm{definition of $q^{\pi_0}$}\}\\
&q^{\pi_0}\,(\just\, p) \plt q^{\pi_0}\, (\just\, p'))\rlap{\hbox to 187 pt{\hfill\qEd}}
\end{array}
\]

\begin{lem}
  The container morphism $f = t^f \lhd q^f$
  is a directed container morphism.
\end{lem}

\proof Auxiliary statements $t^f\, (s \dnp \qfbarzero\, p) =
\tfbarzero\, s \dnpll p$ and $t^f\, (s \dnp \qfbarone\, p) =
\tfbarone\, s \dnplr p$ for law 1, by mutual induction on the $p$s,
showing the cases of the first statement; those of the second one are
symmetric. Case $p = (p_0, \nothing)$.
\[
\begin{array}{cl}
&t^f\, (s \dnp \qfbarzero (p_0 , \nothing))\\
=&\quad\{\textrm{definition of $\qfbarzero$} \}\\
&t^f\, (s \dnp \qfzero\, (\just\, p_0))\\
=&\quad\{\textrm{definition of $t^f$} \}\\
&(\tfbarzero\, (s \dnp q^{f_0}\, (\just\, p_0)),
\tfbarone\, (s \dnp q^{f_0}\, (\just\, p_0)))\\
=&\quad\{\textrm{definition of $\tfbarzero$} \}\\
&((t^{f_0}\, (s \dnp q^{f_0}\, (\just\, p_0)), \lambda p .\, \tfbarone\, ((s \dnp q^{f_0}\, (\just\, p_0)) \dnp q^{f_0}\, (\just\, p))) , 
\tfbarone\, (s \dnp q^{f_0}\, (\just\, p_0)))\\
=&\quad\{\textrm{directed container law 2}\}\\
&((t^{f_0}\, (s \dnp q^{f_0}\, (\just\, p_0), \lambda p .\, \tfbarone\, (s \dnp (q^{f_0}\, (\just\, p_0) \pl' q^{f_0}\, (\just\, p))))) , 
\tfbarone\, (s \dn' q^{f_0}\, (\just\, p_0)))\\
=&\quad\{\textrm{directed container morphism laws 1, 3}\}\\
&((t^{f_0}\, s \dna \just\, p_0, \lambda p .\, \tfbarone\, (s \dnp (q^{f_0}\, (\just\, p_0 \pla  \just\, p))))) , \tfbarone\, (s \dnp q^{f_0}\, (\just\, p_0)))\\
=&\quad\{\textrm{definitions of $\dna$, $\pla$} \}\\
&((t^{f_0}\, s \dnpla p_0, \lambda p .\, \tfbarone\, (s \dnp (q^{f_0}\, (\just\, ( p_0 \plpla  p))))), \tfbarone\, (s \dnp q^{f_0}\, (\just\, p_0)))\\
=&\quad\{\textrm{definition of $\dnpll$} \}\\
&(\tfzero\, s, \lambda p.\, \tfbarone\, (s \dnp \qfzero\, (\just\, p))) 
      \dnpll  (p_0 , \nothing)\\
=&\quad\{\textrm{definition of $\tfbarzero$} \}\\
&\tfbarzero\,s \dnpll (p_0 , \nothing)
\end{array}
\]
Case $p = (p_0 , \just\, p_1)$:
\[
\begin{array}{cl}
&t^f\, (s \dnp \qfbarzero\, (p_0 , \just\, p_1))\\
=&\quad\{\textrm{definition of $\qfbarzero$} \}\\
&t^f\, (s \dnp (q^{f_0}\, (\just\, p_0) \plp \qfbarone\, p_1))\\
=&\quad\{\textrm{directed container law 2 }\}\\
&t^f\, ((s \dnp q^{f_0}\, (\just\, p_0)) \dnp \qfbarone\, p_1)\\
=&\quad\{\textrm{inductive hypothesis for $p_1$}\}\\
&\tfbarone\, (s \dnp q^{f_0}\, (\just\, p_0)) \dnplr p_1\\
=&\quad\{\textrm{definition of $\dnpll$}\}\\
&(\tfzero\, s, \lambda p.\, \tfbarone\, (s \dnp \qfzero\, (\just\, p))) 
      \dnpll  (p_0 , \just\, p_1)\\
=&\quad\{\textrm{definition of $\tfbarzero$}\}\\
&\tfbarzero\, s \dnpll (p_0 , \just\, p_1)
\end{array}
\]

\noindent Directed container morphism law 1. Case $p = \nothing$:
\[
\begin{array}{cl}
&t^f\, (s \dnp q^f\, \nothing) \\
=&\quad\{\textrm{definition of $q^f$} \}\\
&t^f\, (s \dnp \op)\\
=&\quad\{\textrm{directed container law 1}\}\\
&t^f\, s\\
=&\quad\{\textrm{definition of $\dnt$} \}\\
&t^f\, s \dnt \nothing 
\end{array}
\]
Case $p = \just\, (\inl\, p)$:
\[
\begin{array}{cl}
&t^f\, (s \dnp q^f\, (\just\, (\inl\, p))) \\
=&\quad\{\textrm{definition of $q^f$} \}\\
& t^f\, (s \dnp \qfbarzero\, p)\\
=&\quad\{\textrm{aux. statement}\}\\
&\tfbarzero\, s \dnpll p \\
=&\quad\{\textrm{definition of $t^f$, $\dnt$} \}\\
&t^f\, s \dnt \just\, (\inl\, p)
\end{array}
\]
Case $p = \just\, (\inr\, p)$ is symmetric.

\noindent Directed container morphism law 2: 
\[
\begin{array}{cl}
&q^f\,\ot\\
=&\quad\{\textrm{definition of $\ot$} \}\\
&q^f\,\nothing\\
=&\quad\{\textrm{definition of $q^f$} \}\\
&\op
\end{array}
\]

\noindent Auxiliary statements $\qfbarzero\, (p \plpll p') =
\qfbarzero\, p\plp q^f\,(\just\,p')$ and $\qfbarone\, (p \plplr p') =
\qfbarone\, p\plp q^f\,(\just\,p')$ for law 3, by mutual induction on
the $p$s, showing the cases of the first statement. Case $p = (p_0,
\nothing)$, $p' = \inl\,(p_0', \nothing)$:
\[
\begin{array}{cl}
&\qfbarzero\,((p_0, \nothing)\plpll\inl\, (p_0',\nothing))\\
=&\quad\{\textrm{definition of $\plpll$}\}\\
&\qfbarzero\,(p_0 \plpla p_0', \nothing)\\
=&\quad\{\textrm{definition of $\qfbarzero$}\}\\
&\qfzero\,(\just\,(p_0\plpla p_0'))\\
=&\quad\{\textrm{definition of $\pla$}\}\\
&\qfzero\,(\just\,p_0\pla\just\,p_0')\\
=&\quad\{\textrm{directed container morphism law 3}\}\\
&\qfzero\,(\just\,p_0)\plp\qfzero\,(\just\,p_0')\\
=&\quad\{\textrm{definition of $\qfbarzero$}\}\\
&\qfbarzero\,(p_0,\,\nothing)\plp\qfbarzero\,(p_0',\,\nothing)\\
=&\quad\{\textrm{definition of $q^f$}\}\\
&\qfbarzero\,(p_0,\,\nothing)\plp q^f\,(\just\,(\inl\,(p_0',\,\nothing)))
\end{array}
\]
Case $p = (p_0, \nothing)$, $p' = \inl\,(p_0', \just\,p_1')$:
\[
\begin{array}{cl}
&\qfbarzero\,((p_0,\,\nothing)\plpll\inl(p_0',\just\,p_1'))\\
=&\quad\{\textrm{definition of $\plpll$}\}\\
&\qfbarzero\,(p_0 \plpla p_0', \just\,p_1')\\
=&\quad\{\textrm{definition of $\qfbarzero$}\}\\
&\qfzero\,(\just\,(p_0\plpla p_0')) \plp \qfbarone\,p_1'\\
=&\quad\{\textrm{definition of $\pla$}\}\\
&\qfzero\,(\just\,p_0\pla\just\,p_0') \plp \qfbarone\,p_1'\\
=&\quad\{\textrm{directed container morphism law 3}\}\\
&(\qfzero\,(\just\,p_0)\plp\qfzero\,(\just\,p_0')) \plp \qfbarone\,p_1'\\
=&\quad\{\textrm{directed container law 5}\}\\
&\qfzero\,(\just\,p_0)\plp(\qfzero\,(\just\,p_0') \plp \qfbarone\,p_1')\\
=&\quad\{\textrm{definition of $\qfbarzero$}\}\\
&\qfbarzero\,(p_0,\,\nothing)\plp\qfbarzero\,(p_0',\,\just\,p_1')\\
=&\quad\{\textrm{definition of $q^f$}\}\\
&\qfbarzero\,(p_0,\,\nothing)\plp q^f\,(\just\,(\inl\,(p_0',\,\just\,p_1')))
\end{array}
\]
Case $p = (p_0, \nothing)$, $p' = \inr\,p'$:
\[
\begin{array}{cl}
&\qfbarzero\,((p_0,\,\nothing)\plpll\inr\,p')\\
=&\quad\{\textrm{definition of $\plpll$}\}\\
&\qfbarzero\,(p_0,\,\just\,p')\\
=&\quad\{\textrm{definition of $\qfbarzero$}\}\\
&\qfzero\,(\just\,p_0)\plp\qfbarone\,p'\\
=&\quad\{\textrm{definitions of $\qfbarzero$ and $q^f$}\}\\
&\qfbarzero\,(p_0,\,\nothing)\plp q^f\,(\just\,(\inr\,p'))
\end{array}
\]
Case $p = (p_0, \just\, p_1)$:
\[
\begin{array}{cl}
&\qfbarzero\,((p_0,\,\just\,p_1) \plpll p')\\
=&\quad\{\textrm{definition of $\plpll$}\}\\
&\qfbarzero\,(p_0,\,\just\,(p_1 \plplr p'))\\
=&\quad\{\textrm{definition of $\qfbarzero$}\}\\
&\qfzero\,(\just\,p_0)\plp \qfbarone\,(p_1\plplr p')\\
=&\quad\{\textrm{inductive hypothesis for $p_1$}\}\\
&\qfzero\,(\just\,p_0)\plp(\qfbarone\,p_1\plp q^f\,(\just\,p'))\\
=&\quad\{\textrm{directed container law 5}\}\\
&(\qfzero\,(\just\,p_0)\plp\qfbarone\,p_1)\plp q^f\,(\just\,p')\\
=&\quad\{\textrm{definition of $\qfbarzero$}\}\\
&\qfbarzero\,(p_0,\,\just\,p_1)\plp q^f\,(\just\,p')
\end{array}
\]

\noindent Directed container law 3. Case $p = \nothing$:
\[
\begin{array}{cl}
&q^f\,(\nothing \plt p')\\
=&\quad\{\textrm{definition of $\plt$ }\}\\
&q^f\, p'\\
=&\quad\{\textrm{directed container law 4}\}\\
&\op \plp q^f\, p'\\
=&\quad\{\textrm{definition of $q^f$} \}\\
&q^f\, \nothing \plp q^f\, p'\\
\end{array}
\]
Case $p= \just\, p$, $p' = \nothing$:
\[
\begin{array}{cl}
&q^f\,(\just\, p \plt \nothing)\\
=&\quad\{\textrm{definition of $\plt$} \}\\
&q^f\, (\just\, p)\\
=&\quad\{\textrm{directed container law 3}\}\\
&q^f\, (\just\, p) \plp \op\\
=&\quad\{\textrm{definition of $q^f$} \}\\
&q^f\, (\just\, p) \plp q^f\, \nothing
\end{array}
\]
Case $p = \just\, (\inl\, p)$, $p' = \just\, p'$:
\[
\begin{array}{cl}
&q^f\,(\just\, (\inl\, p) \plt \just\, p')\\
=&\quad\{\textrm{definition of $\plt$ }\}\\
&q^f\,(\just\, (\inl\, p \plpl p'))\\
=&\quad\{\textrm{definition of $\plpl$ }\}\\
&q^f\,(\just\, (\inl\, (p \plpla p')))\\
=&\quad\{\textrm{definition of $q^f$} \}\\
&\qfbarzero\, (p \plt p')\\
=&\quad\{\textrm{first aux. statement}\}\\
& \qfbarzero\, p\plp q^f\,(\just\,p')\\
=&\quad\{\textrm{definition of $\qfbarzero$ }\}\\
&q^f\, (\just\, (\inl\, p)) \plp q^f\, (\just\,p')
\end{array}
\]
Case $p = \just\, (\inr\, p)$, $p' = \just\, p'$ is symmetric.
\qed

\begin{lem} 
The product triangles $\eqref{diag:products}$ commute, i.e., $\pi_0
\comp f = f_0$ and $\pi_1 \comp f = f_1$.
\end{lem}
\proof We verify only the left triangle $\pi_0 \comp f = f_0$.
The right triangle is symmetric.

\noindent Statement for shapes:
\[
\begin{array}{cl}
&t^{\pi_0}\,(t^f\,s)\\
=&\quad\{\textrm{definition of $t^{\pi_0}$} \}\\
&\fst\, (\fst\,  (t^f\, s)) \\
=&\quad\{\textrm{definition of $t^f$} \}\\
&\fst\,  (\tfbarzero\, s)\\
=&\quad\{\textrm{definition of $\tfbarzero$} \}\\
&t^{f_0}\,s
\end{array}
\]

\noindent Statement for positions.
Case $p = \nothing$:
\[
\begin{array}{cl}
&q^f(q^{\pi_0}\,\nothing)\\
=&\quad\{\textrm{definition of $q^{\pi_0}$} \}\\
&q^f\nothing\\
=&\quad\{\textrm{definition of $q^f$} \}\\
&\op\\
=&\quad\{\textrm{directed container morphism law 2} \}\\
&q^{f_0}\,\o_0\\
=&\quad\{\textrm{definition of $\o_0$} \}\\
&q^{f_0}\,\nothing
\end{array}
\]
Case $p = \just\,p$:
\[
\begin{array}{cl}
&q^f(q^{\pi_0}\,(\just\,p))\\
=&\quad\{\textrm{definition of $q^{\pi_0}$} \}\\
&q^f\,(\just\,(\inl\,(p,\,\nothing)))\\
=&\quad\{\textrm{definition of $q^f$} \}\\
&\qfbarzero\, (p, \nothing)\\
=&\quad\{\textrm{definition of $\qfbarzero$} \}\\
&q^{f_0}\,(\just\,p)\rlap{\hbox to 217 pt{\hfill\qEd}} 
\end{array}
\]
\newpage

\begin{lem}
  The directed container morphism $f = t^f \lhd q^f$
  is unique, i.e., if there is a directed container morphism $h =
  t^h\lhd q^h : E' \to E$ such that $\pi_0 \comp h = f_0$ and $\pi_1
  \comp h = f_1$, then $f = h$.
\end{lem}

\proof Auxiliary statements $\tfbarzero\, s = \fst\, (t^h\,
s)$ and $\tfbarone\, s = \snd\, (t^h\, s)$ for shapes, by mutual
coinduction, showing only the case of the first statement.
\[
\begin{array}{cl}
&\tfbarzero \,s\\
=&\quad\{\textrm{definition of $\tfbarone$}\}\\
&(t^{f_0}\,s,\lambda p_0.\,\tfbarone\,(s\dnp\,q^{f_0}\,(\just\,p_0)))\\
=&\quad\{\textrm{assumption}\}\\
&(t^{\pi_0}\,(t^h\,s), \lambda p_0.\,\tfbarone\,(s\dnp\,q^{f_0}\,(\just\,p_0)))\\
=&\quad\{\textrm{coinductive hypothesis}\}\\
&(t^{\pi_0}\,(t^h\,s),\lambda p_0.\,\snd\,(t^h\,(s\dnp\,q^h\,(q^{\pi_0}\,(\just\,p_0))))\\
=&\quad\{\textrm{directed container morphism law 1}\}\\
&(t^{\pi_0}\,(t^h\,s),\lambda p_0.\,\snd\,(t^h\,s \dnt\, q^{\pi_0}\,(\just\,p_0)))\\
=&\quad\{\textrm{definition of $q^{\pi_0}$}\}\\
&(t^{\pi_0}\,(t^h\,s),\lambda p_0.\,\snd\,(t^h\,s\dnt\,\just\,(\inl\,(p_0,\nothing))))\\
=&\quad\{\textrm{definition of $\dnt$} \}\\
&(t^{\pi_0}\,(t^h\,s),\lambda p_0.\,\snd\,(t^h\,s\dntpl\inl\,(p_0,\nothing)))\\
=&\quad\{\textrm{definition of $\dntpl$} \}\\
&(t^{\pi_0}\,(t^h\,s),\lambda p_0.\,\snd\,(\fst\,(t^h\,s)\dnpll (p_0,\nothing)))\\
=&\quad\{\textrm{definition of $\dnpll$} \}\\
&(t^{\pi_0}\,(t^h\,s),\lambda p_0.\,\snd\,(\fst\,(t^h\,s))\,p_0)\\
=&\quad\{\textrm{definition of $t^{\pi_0}$}\}\\
&\fst\, (t^h\,s)
\end{array}
\]

\noindent Statement for shapes, i.e., $t^f = t^h$:
\[
\begin{array}{cl}
& t^f\, s\\
=&\quad\{\textrm{definition of $t^f$}\}\\
& (\tfbarzero\, s, \tfbarone\, s)\\
=&\quad\{\textrm{aux. statements}\}\\
& t^h\, s
\end{array}
\]

\noindent Auxiliary statements $\qfbarzero\, p = q^h\, (\just\,
(\inl\, p))$ and $\qfbarone\, p = q^h\, (\just\, (\inr\, p))$ for
positions, by mutual induction on the $p$s, showing the cases of the
first statement. Case $p = (p_0,\nothing)$:
\[
\begin{array}{cl}
&\qfbarzero\,(p_0,\nothing)))\\
=&\quad\{\textrm{definition of $\qfbarzero$}\}\\
&q^{f_0}\,(\just\,p_0)\\
=&\quad\{\textrm{assumption}\}\\
&q^h\,(q^{\pi_0}\,(\just\,p_0))\\
=&\quad\{\textrm{definition of $q^{\pi_0}$}\}\\
&q^h\,(\just\,(\inl\,(p_0,\nothing)))\\
\end{array}
\]
Case $p = (p_0,\just\,p_1)$:
\[
\begin{array}{cl}
& \qfbarzero\, (p_0,\just\,p_1)))\\
=&\quad\{\textrm{definition of $\qfbarzero$}\}\\
&q^{f_0}\,(\just\,p_0)\plp \qfbarone\, p_1\\
=&\quad\{\textrm{assumption}\}\\
&q^h\,(q^{\pi_0}\, (\just\,p_0)))\plp \qfbarone\, p_1\\
=&\quad\{\textrm{inductive hypothesis for $p_1$}\}\\
&q^h\,(q^{\pi_0}\, (\just\,p_0)))\plp q^h\,(\just\,(\inr\,p_1))\\
=&\quad\{\textrm{definition of $q^{\pi_0}$}\}\\
&q^h\,(\just\,(\inl\,(p_0,\nothing)))\plp q^h\,(\just\,(\inr\,p_1))\\
=&\quad\{\textrm{directed container morphism law 3}\}\\
&q^h\,(\just\,(\inl\,(p_0,\nothing))\plt \just\,(\inr\,p_1))\\
=&\quad\{\textrm{definition of $\plt$}\}\\
&q^h\,(\just\,(\inl\,(p_0,\nothing)\plpl\inr\,p_1))\\
=&\quad\{\textrm{definition of $\plpl$}\}\\
&q^h\,(\just\,(\inl\, ((p_0,\nothing)\plpll\inr\,p_1)))\\
=&\quad\{\textrm{definition of $\plpll$}\}\\
&q^h\,(\just\,(\inl\,(p_0,\just\,p_1)))
\end{array}
\]

\noindent Statement for positions, i.e., $q^f = q^h$. Case $p =
\nothing$:
\[
\begin{array}{cl}
&q^f\,\nothing\\
=&\quad\{\textrm{definition of $q^f$}\}\\
&\op \\
=&\quad\{\textrm{directed container morphism law 2}\}\\
&q^h\, \ot\\
=&\quad\{\textrm{definition of $\ot$}\}\\
&q^h\,\nothing
\end{array}
\]
Case $p = \just\, (\inl\, p)$:
\[
\begin{array}{cl}
&q^f\,(\just\,(\inl\,p)\\
=&\quad\{\textrm{definition of $q^f$}\}\\
&\qfbarzero\, p\\
=&\quad\{\textrm{first aux. statement}\}\\
&q^h\,(\just\,(\inl\,p))
\end{array}
\]
Case $p = \just\, (\inr\, p)$ is symmetric.
\qed\newpage


\subsection*{Proof of Proposition~\ref{prop:cofree}}\hfill

We must prove that $E = (S \lhd P, \dn, \o, \pl)$ is a cofree directed
container on the container $C_0 = S_0 \lhd P_0$.

\begin{lem}
The data $\dn$, $\o$, $\pl$ provide a directed container structure on
the container $C = S \lhd P$.
\end{lem}

\proof
Directed container law 1:
\[
\begin{array}{cl}
&(s,v ) \dn \o\, \ia{s,v}\\
=&\quad\{\textrm{definition of $\o$}\}\\
&(s,v) \dn (\inl\, \zt)\\
=&\quad\{\textrm{definition of $\dn$}\}\\
&(s, v)
\end{array}
\]

\noindent Directed container law 2 by induction on $p$. Case $p=\inl\, \zt$:
\[
\begin{array}{cl}
&(s, v) \dn (\inl\, \zt \pl p')\\
=&\quad\{\textrm{definition of $\pl$}\}\\
&(s, v) \dn p'\\
=&\quad\{\textrm{definition of $\dn$}\}\\
&((s, v) \dn \inl\, \zt) \dn p'
\end{array}
\]
Case $p=\inr\, (p , p')$:
\[
\begin{array}{cl}
&(s,v) \dn (\inr\, (p , p') \pl p'')\\
=&\quad\{\textrm{definition of $\pl$}\}\\
&(s,v) \dn (\inr\, (p , p'\pl p''))\\
=&\quad\{\textrm{definition of $\dn$}\}\\
&v\, p \dn (p' \pl p'')\\
=&\quad\{\textrm{inductive hypothesis for $p'$}\}\\
&(v\, p \dn p') \dn p''\\
=&\quad\{\textrm{definition of $\dn$}\}\\
&((s,v) \dn \inr\, (p , p')) \dn p''
\end{array}
\]

\noindent Directed container law 3 by induction on $p$. Case $p=\inl\, \zt$: 
\[
\begin{array}{cl}
&\inl\, \zt \pl \o\, \ia{(s,v) \dn \inl\, \zt}\\
=&\quad\{\textrm{definitions of $\pl$, $\dn$}\}\\
&\o\, \ia{s,v}\\
=&\quad\{\textrm{definition of $\o$}\}\\
&\inl\, \zt
\end{array}
\]
Case $p=\inr\, (p , p')$:
\[
\begin{array}{cl}
&\inr\, (p , p') \pl \o\, \ia{(s, v) \dn \inr\, (p , p')}\\
=&\quad\{\textrm{definitions of $\pl$, $\dn$}\}\\
&\inr\, (p , p' \pl \o\, \ia{v\, p \dn p'})\\
=&\quad\{\textrm{inductive hypothesis for $p'$}\}\\
&\inr\, (p , p')
\end{array}
\]

\noindent Directed container law 4.
\[
\begin{array}{cl}
&\o\, \ia{s, v} \pl p\\
=&\quad\{\textrm{definition of $\o$} \}\\
&\inl\, \zt \pl p\\
=&\quad\{\textrm{definition of $\pl$}\}\\
&p
\end{array}
\]

\noindent Directed container law 5 by induction on $p$. Case $p=\inl\, \zt$: 
\[
\begin{array}{cl}
&(\inl\, \zt \pl p') \pl p''\\
=&\quad\{\textrm{definition of $\pl$}\}\\
&p' \pl p''\\
=&\quad\{\textrm{definition of $\pl$} \}\\
&\inl\, \zt \pl (p' \pl p'')
\end{array}
\]
Case $p=\inr\, (p , p')$:
\[
\begin{array}{cl}
&(\inr\, (p , p') \pl p'') \pl p'''\\
=&\quad\{\textrm{definition of $\pl$}\}\\
&\inr\, (p , p' \pl p'') \pl p'''\\
=&\quad\{\textrm{definition of $\pl$}\}\\
&\inr\, (p , (p' \pl p'') \pl p''')\\
=&\quad\{\textrm{inductive hypothesis for $p'$} \}\\
&\inr\, (p , p' \pl (p'' \pl p'''))\\
=&\quad\{\textrm{definition of $\pl$}\}\\
&\inr\, (p , p') \pl (p'' \pl p''')\rlap{\hbox to 179 pt{\hfill\qEd}} 
\end{array} 
\]\medskip

\noindent That the directed container $E = (S \lhd P, \dn, \o, \pl)$ is cofree
on the container $C_0 = S_0 \lhd P_0$ can be shown either
directly or by proving that it interprets into a cofree comonad on
$\csem{C_0}$.  In the following, we illustrate the first route. This
involves a fair amount of straightforward, but tedious inductive and
coinductive reasoning in the lemmas below.

For the directed container $E$ to be cofree on the container $C_0$,
there must be a container morphism $\pii = t^{\pii}\lhd q^{\pii} : S
\lhd P \to S_0 \lhd P_0$.  This is defined by
\begin{itemize}
\item $t^{\pii} : S \to S_0$\\
      $t^{\pii}\,(s, v) = s$
\item $q^{\pii} : \Pi \ia{(s,  v) : S}.\, 
                       P_0\, s \to P\,(s, v)$\\
      $q^{\pii}\,p = \inr\,(p, \inl\,\zt)$
\end{itemize}
The universal property of cofreeness states that, for any other
directed container $E' = (S' \lhd P', \dnp, \op, \plp)$ and container
morphism $f_0 = t^{f_0} \lhd q^{f_0} : S' \lhd P' \to S_0 \lhd P_0 $, there must
exist a unique directed container morphism $f = t^f \lhd q^f : E' \to
E$ such that the following triangle commutes:
\[
\tag{$\ddag$}
\label{diag:cofreeness}
\begin{gathered}
\xymatrix@C=3.5em@R=3em{
S' \lhd P'\ar[d]_{f} \ar[rd]^{f_0} & \\
S \lhd P\ar[r]^{\pii}&S_0 \lhd P_0
}
\end{gathered}
\]
We claim that this directed container morphism $f$ is given by
\begin{itemize}
\item $t^f : S' \to S$\\ 
      (by corecursion) \\
      $t^f\,s = (t^{f_0}\,s, \lambda p.\,t^f\,(s\dnp q^{f_0}\,p))$
\item 
      $q^f : \Pi \ia{s : S'}.\,P\,(t^{f_0}\,s, \lambda p.\,t^f\,(s\dnp q^{f_0}\,p)) \to P'\,s$\\
      (by recursion) \\
      $q^f\, (\inl\,\zt) = \op$\\
      $q^f\, (\inr\,(p, p')) = q^{f_0} p \plp q^f\,p$
\end{itemize}
and prove it with the lemmas below.

\begin{lem}
  The container morphism $f$ is a directed container morphism.
\end{lem}

\proof Directed container morphism law 1 by induction on $p$. Case $p
= \inl\,\zt$:
\[
\begin{array}{cl}
&t^f\, (s \dnp q^f\, (\inl\,\zt))\\
=&\quad\{\textrm{definition of $q^f$}\}\\
&t^f\, (s \dnp \op)\\
=&\quad\{\textrm{directed container law 1}\}\\
&t^f\,s\\
=&\quad\{\textrm{definition of $\dn$}\}\\
&t^f\,s\dn \inl\, \zt\\
\end{array}
\]
Case $p = \inr\,(p, p')$:
\[
\begin{array}{cl}
&t^f\, (s\dnp q^f\,(\inr\,(p, p')))\\
=&\quad\{\textrm{definition of $q^f$}\}\\
&t^f\, (s\dnp (q^{f_0}\, p \plp q^f\, p))\\
=&\quad\{\textrm{directed container law 2}\}\\
&t^f\, ((s\dnp q^{f_0}\, p) \dnp q^f\, p')\\
=&\quad\{\textrm{inductive hypothesis for $p'$}\}\\
&t^f\, (s\dnp q^{f_0}\,p)\dn p'\\
=&\quad\{\textrm{definition of $\dn$}\}\\
&(t^{f_0}\,s, \lambda p.\,t^f\, (s\dnp q^{f_0}\,p))\dn \inr\,(p, p')\\
=&\quad\{\textrm{definition of $t^f$}\}\\
&t^f\,s\dn \inr\,(p, p')
\end{array}
\]

\noindent
Directed container morphism law 2.
\[
\begin{array}{cl}
&q^f\,\o\\
=&\quad\{\textrm{definition of $\o$}\}\\
&q^f\,(\inl\, \zt)\\
=&\quad\{\textrm{definition of $q^f$}\}\\
&\op
\end{array}
\]

\noindent
Directed container morphism law 3 by induction on $p$. Case $p =
\inl\,\zt$:
\[
\begin{array}{cl}
&q^f\,(\inl\, \zt \pl p')\\
=&\quad\{\textrm{definition of $\pl$}\}\\
&q^f p'\\
=&\quad\{\textrm{directed container law 4}\} \\
&\op \plp q^f\,p'\\
=&\quad\{\textrm{definition of $q^f$}\}\\
&q^f\,(\inl\,\zt)\plp q^f\,p'
\end{array}
\]
Case $p = \inr\, (p, p')$:
\[
\begin{array}{cl}
&q^f\,(\inr\, (p, p')\pl p'')\\
=&\quad\{\textrm{definition of $\pl$}\}\\
&q^f\,(\inr\,(p, p'\plp p''))\\
=&\quad\{\textrm{definition of $q^f$}\}\\
&q^{f_0}\,p \plp q^f(p' \plp p'')\\
=&\quad\{\textrm{inductive hypothesis for $p'$}\}\\ 
&q^{f_0}\,p\plp (q^f\,p' \plp q^f\,p'')\\
=&\quad\{\textrm{directed container law 5}\}\\
&(q^{f_0}\,p \plp q^f\,p') \plp q^f\,p''\\
=&\quad\{\textrm{definition of $q^f$}\}\\
&q^f\,(\inr\, (p, p')) \plp q^f\,p''\rlap{\hbox to 179 pt{\hfill\qEd}} 
\end{array}
\]

\begin{lem}
The cofreeness triangle \eqref{diag:cofreeness} commutes, i.e., $\pii
\comp f = f_0$.
\end{lem}
\proof
Statement for shapes, i.e., $t^{\pii}\comp t^f = t^{f_0}$: 
\[
\begin{array}{cl}
&t^{\pii}\,(t^f\,s)\\
=&\quad\{\textrm{definition of $t^f$}\}\\
&t^{\pii}\,(t^{f_0}\,s,\lambda p.\,t^f\,(s\dnp q^{f_0}\,p))\\
=&\quad\{\textrm{definition of $t^{\pii}$}\}\\
&t^{f_0}\,s
\end{array}
\]
Statement for positions, i.e., $q^f \comp q^{\pii} = q^{f_0}$:
\[
\begin{array}{cl}
&q^f\,(q^{\pii}\,p)\\
=&\quad\{\textrm{definition of $q^{\pii}$}\}\\
&q^f\,(\inr\,(p, \inl\,\zt))\\
=&\quad\{\textrm{definition of $q^f$}\}\\
&q^{f_0}\,p\plp q^f\,(\inl\,\zt)\\
=&\quad\{\textrm{definition of $q^f$}\}\\
&q^{f_0}\,p\plp \op\\
=&\quad\{\textrm{directed container law 3}\}\\
&q^{f_0}\,p\rlap{\hbox to 254 pt{\hfill\qEd}} 
\end{array}
\]\newpage

\begin{lem}
  The directed container morphism $f$ is unique, i.e., if there is a
  directed container morphism $h = t^h \lhd q^h: E' \to E$ such that
  $\pii \comp h = f_0$, then $f = h$.
\end{lem}

\proof
Statement for shapes, i.e., $t^f = t^h$, by coinduction.
\[
\begin{array}{cl}
&t^f\,s\\
=&\quad\{\textrm{definition of $t^f$}\}\\
&(t^{f_0}\,s, \lambda p.\,t^f\,(s\dnp q^{f_0}\, p))\\
=&\quad\{\textrm{coinductive hypothesis}\}\\
&(t^{f_0}\,s, \lambda p.\,t^h\,(s\dnp q^{f_0}\, p))\\
=&\quad\{\textrm{assumption, i.e., $t^{\pii}\comp t^h = t^{f_0}$ and $q^h\comp q^{\pii} = q^{f_0}$}\}\\
&(t^\pii\,(t^h\,s), \lambda p.\,t^h\,(s\dnp q^h\, (q^{\pii}\, p)))\\
=&\quad\{\textrm{directed container morphism law 1}\}\\
&(t^\pii\,(t^h\,s), \lambda p.\,t^h\,s\dn q^{\pii} p)\\
=&\quad\{\textrm{definitions of $t^{\pii}$, $q^{\pii}$}\}\\
&(\fst\,(t^h\,s), \lambda p.\,t^h\,s\dn \inr\, (p, \inl\,\zt))\\
=&\quad\{\textrm{definition of $\dn$}\}\\
&(\fst\,(t^h\,s), \lambda p.\,\snd\,(t^h\,s)\,p \dn \inl\,\zt)\\
=&\quad\{\textrm{definition of $\dn$}\}\\
&t^h\,s
\end{array}
\]

\noindent Statement for positions, i.e., $q^f = q^h$, by induction on
position $p$. Case $p = \inl\,\zt$:
\[
\begin{array}{cl}
&q^f\,(\inl\,\zt)\\
=&\quad\{\textrm{definition of $q^f$}\}\\
&\o'\\
=&\quad\{\textrm{directed container morphism law 2}\}\\
&q^h\,\o\\
=&\quad\{\textrm{definition of $\o$}\}\\
&q^h\,(\inl\,\zt)
\end{array}
\]
Case $p = \inr\,(p, p')$:
\[
\begin{array}{cl}
&q^f\,(\inr\,(p, p'))\\
=&\quad\{\textrm{definition of $q^f$}\}\\
&q^{f_0}\,p\plp q^f\,p'\\
=&\quad\{\textrm{inductive hypothesis for $p'$}\}\\
&q^{f_0}\,p\plp q^h\,p'\\
=&\quad\{\textrm{assumption for positions, i.e., $q^h \comp q^{\pii} = q^{f_0}$}\}\\
&q^h\, (q^{\pii}\,p) \plp q^h\,p'\\
=&\quad\{\textrm{directed container morphism law 3}\}\\
&q^h\, (q^{\pii}\,p \pl p')\\
=&\quad\{\textrm{definition of $q^{\pii}$}\}\\
&q^h\, (\inr\,(p,\inl\,\zt)\pl p')\\
=&\quad\{\textrm{definition of $\pl$}\}\\
&q^h\, (\inr\,(p,\inl\,\zt \pl p')\\
=&\quad\{\textrm{definition of $\pl$}\}\\
&q^h\, (\inr\,(p, p'))\rlap{\hbox to 258 pt{\hfill\qEd}} 
\end{array}
\]
\vspace{-20 pt}

\end{document}